\documentclass[useAMS,usenatbib]{mnras}
\usepackage{graphicx}
\usepackage{amsmath}
\usepackage{amssymb}
\usepackage{color}

\def \etal {et~al.~}
\def \chisq  {\ifmmode  \chi^2   \else  $\chi^2$  \fi}  
\def \spose#1{\hbox  to 0pt{#1\hss}}  
\def \lta{\mathrel{\spose{\lower 3pt\hbox{$\sim$}}\raise  2.0pt\hbox{$<$}}}
\def \gta{\mathrel{\spose{\lower  3pt\hbox{$\sim$}}\raise 2.0pt\hbox{$>$}}}

\def \kms {\ifmmode  \,\rm km\,s^{-1} \else $\,\rm km\,s^{-1}  $ \fi }
\def \kpc {\ifmmode  {\rm~kpc}  \else ${\rm~kpc}$\fi}  
\def \pc {\ifmmode  {\rm~pc}  \else ${\rm~pc}$ \fi  }  
\def \Gyr {\ifmmode  {\rm~Gyr}  \else ${\rm~Gyr}$\fi}
\def \Msun {\ifmmode {\rm M}_{\odot} \else ${\rm M}_{\odot}$ \fi} 
\def \Lsun {\ifmmode L_{\odot} \else $L_{\odot}$ \fi} 
\def \Rsun {\ifmmode R_{\odot} \else $R_{\odot}$ \fi} 
\def \Msunpyr {\ifmmode M_{\odot}{\rm~yr}^{-1} \else $M_{\odot}{\rm~yr}^{-1}$ \fi} 
\def \hMsun {\ifmmode h^{-1}\,\rm M_{\odot} \else $h^{-1}\,\rm M_{\odot}$ \fi}
\def \hMpc {\ifmmode  {h^{-1}\rm Mpc}  \else ${h^{-1}\rm Mpc}$ \fi  }  
\def \Mpch {\ifmmode  {h^{-1}\rm Mpc}  \else ${h^{-1}\rm Mpc}$ \fi  }

\def \LCDM {\ifmmode \Lambda{\rm CDM} \else $\Lambda{\rm CDM}$ \fi}
\def \sig8 {\ifmmode \sigma_8 \else $\sigma_8$ \fi} 
\def \OmegaM {\ifmmode \Omega_{\rm M} \else $\Omega_{\rm M}$ \fi} 
\def \OmegaL {\ifmmode \Omega_{\rm \Lambda} \else $\Omega_{\rm \Lambda}$\fi} 
\def \Deltavir {\ifmmode \Delta_{\rm vir} \else $\Delta_{\rm vir}$ \fi}
\def \rhocrit {\ifmmode \rho_{\rm crit} \else $\rho_{\rm crit}$ \fi}
\def \rhou {\ifmmode \rho_{\rm u} \else $\rho_{\rm u}$ \fi}
\def \zc {\ifmmode z_{\rm c} \else $z_{\rm c}$ \fi}

\def\LCDM{$\Lambda$CDM }
\def \LCDM {\ifmmode \Lambda{\rm CDM} \else $\Lambda{\rm CDM}$ \fi}
\def \sig8 {\ifmmode \sigma_8 \else $\sigma_8$ \fi} 
\def \Omegam {\ifmmode \Omega_{\rm m} \else $\Omega_{\rm m}$ \fi} 
\def \Omegab {\ifmmode \Omega_{\rm b} \else $\Omega_{\rm b}$ \fi} 
\def \Omegar {\ifmmode \Omega_{\rm r} \else $\Omega_{\rm r}$ \fi} 
\def \fbar {\ifmmode f_{\rm bar} \else $f_{\rm bar}$ \fi} 
\def \OmegaL {\ifmmode \Omega_{\rm \Lambda} \else $\Omega_{\rm \Lambda}$\fi} 
\def \Deltavir {\ifmmode \Delta_{\rm vir} \else $\Delta_{\rm vir}$ \fi}
\def \rhocrit {\ifmmode \rho_{\rm crit} \else $\rho_{\rm crit}$ \fi}

\def \rhos {\ifmmode \rho_{\rm s} \else $\rho_{\rm s}$ \fi} 
\def \rs {\ifmmode r_{\rm s} \else $r_{\rm s}$ \fi} 
\def \cvir {\ifmmode c_{\rm vir} \else $c_{\rm vir}$ \fi} 
\def \Rvir {\ifmmode r_{\rm vir} \else $R_{\rm vir}$ \fi}
\def \Vvir {\ifmmode V_{\rm  vir} \else  $V_{\rm vir}$  \fi} 
\def \Mvir {\ifmmode M_{\rm  vir} \else $M_{\rm  vir}$ \fi}  
\def \Nvir {\ifmmode N_{\rm  vir} \else $N_{\rm  vir}$ \fi}  
\def \Jvir {\ifmmode J_{\rm vir} \else $J_{\rm vir}$ \fi} 
\def \Evir {\ifmmode E_{\rm vir} \else $E_{\rm vir}$ \fi} 
\def \vvir {\ifmmode v_{\rm vir} \else $v_{\rm vir}$ \fi} 
\def \lam {\ifmmode \lambda  \else $\lambda$ \fi} 
\def \lamp {\ifmmode \lambda^{\prime} \else $\lambda^{\prime}$  \fi} 
\def \Vmax {\ifmmode V_{\rm  max} \else  $V_{\rm max}$  \fi} 
\def \Mdm {\ifmmode M_{\rm  dm} \else $M_{\rm  dm}$ \fi}

\def \Mgas {\ifmmode M_{\rm gas} \else $M_{\rm gas}$ \fi} 
\def \Mcg {\ifmmode M_{\rm cg} \else $M_{\rm cg}$\fi} 
\def \Mhg {\ifmmode M_{\rm hg} \else $M_{\rm hg}$ \fi} 
\def \Mdisc {\ifmmode M_{\rm disc} \else $M_{\rm disc}$ \fi} 
\def \Md {\ifmmode M_{\rm d} \else $M_{\rm d}$ \fi} 
\def \Mda {\ifmmode M_{\rm d,0\%} \else $M_{\rm d,0\%}$ \fi} 
\def \Mdb {\ifmmode M_{\rm d,20\%} \else $M_{\rm d,20\%}$ \fi} 
\def \Mdc {\ifmmode M_{\rm d,40\%} \else $M_{\rm d,40\%}$ \fi} 
\def \md {\ifmmode m_{\rm d} \else $m_{\rm d}$ \fi} 
\def \Mb {\ifmmode M_{\rm b} \else $M_{\rm b}$ \fi} 
\def \Mbh {\ifmmode M_{\rm b,pri} \else $M_{\rm b,pri}$ \fi} 
\def \Mbs {\ifmmode M_{\rm b,sat} \else $M_{\rm b,sat}$ \fi} 
\def \zo {\ifmmode z_{0} \else $z_{0}$ \fi} 
\def \rd {\ifmmode r_{\rm d} \else $r_{\rm d}$ \fi}
\def \rg {\ifmmode r_{\rm g} \else $r_{\rm g}$ \fi}
\def \rb {\ifmmode r_{\rm b} \else $r_{\rm b}$\fi}
\def \rs {\ifmmode r_{\rm s} \else $r_{\rm s}$\fi}
\def \rc {\ifmmode r_{\rm c} \else $r_{\rm c}$\fi}
\def \rvir {\ifmmode r_{\rm vir} \else $r_{\rm vir}$\fi}
\def \rbh {\ifmmode r_{\rm b,pri} \else $r_{\rm b,pri}$ \fi} 
\def \rbs {\ifmmode r_{\rm b,sat} \else $r_{\rm b,sat}$ \fi}

\title[Edge of galaxy formation I]{The edge of galaxy formation I: formation and evolution of MW-satellites
  analogues before accretion} 
\author[A.V. Macci\`o et al.]{Andrea V. Macci\`o$^{1,2}$\thanks{maccio@nyu.edu}, Jonas Frings$^{2,3}$, Tobias Buck$^2$,
  Camilla Penzo$^{4,2}$, 
  \newauthor{Aaron A. Dutton$^1$, Marvin Blank$^{1,5}$, Aura Obreja$^{1,6}$}
  \\
 $^{1}$New York University Abu Dhabi, PO Box 129188, Saadiyat Island, Abu Dhabi, United Arab Emirates\\
 $^{2}$Max Planck Institute f\"{u}r Astronomie, K\"{o}nigstuhl 17, D-69117 Heidelberg, Germany\\
 $^{3}$Astronomisches Recheninstitut, Zentrum f\"ur Astronomie der Universit\"at Heidelberg, Philosophenweg 12, D-69120 Heidelberg, Germany\\
 $^{4}$Laboratoire\,Univers\,et\,Th\'eories,\,UMR\,8102\,CNRS,\,Observatoire\,de\,Paris,\,Universit\'e\,Paris Diderot,\,5 Place\,Jules\,Janssen,\,92190\,Meudon,\,France\\
 $^{5}$Institut f\"{u}r Theoretische Physik und Astrophysik, Christian-Albrechts-Universit\"at zu Kiel, Leibnizstr. 15, D-24118 Kiel, Germany\\
  $^{6}$Universit\"ats-Sternwarte, Ludwig-Maximilians-Universit\"at M\"unchen, Scheinerstr. 1, D-81679 M\"unchen,
  Germany
  }

\begin{document}

\maketitle

\label{firstpage}

\begin{abstract}
The satellites of the Milky Way and Andromeda represent the smallest
galaxies we can observe  in our Universe. In this series of papers we
aim to shed light on their formation and evolution using cosmological
hydrodynamical simulations. In this first paper we focus on the galaxy
properties before accretion, by simulating twenty seven haloes with masses
between $5\times 10^8$  and $10^{10}\Msun$. Out of this set  nineteen
haloes successfully form stars, while eight remain dark.  The
simulated galaxies match quite well present day observed scaling
relations  between stellar mass, size and metallicity, showing that
such relations are in place before accretion.  Our galaxies show a
large variety of star formation histories, from extended star
formation periods to single bursts. As in more massive galaxies, large
star formation bursts are connected with major mergers events, which
greatly contribute to the overall stellar mass build up. The intrinsic
stochasticity of mergers  induces a large scatter in the stellar mass
halo mass relation, up to two orders of magnitude. 
Despite the bursty star formation history, on these
mass scales baryons are very ineffective in modifying the dark matter
profiles, and galaxies with a stellar mass below $\approx 10^6 \Msun$ retain their cuspy central dark matter
distribution, very similar to results from pure N-body simulations. 

\end{abstract}


\section{Introduction}

In a universe dominated by Dark Matter and Dark Energy, galaxy
formation is a complicated mixture of dark matter assembly, gas infall
and secular evolution.  In the current model for structure  formation
\cite[e.g.][]{White1978,Blumenthal1984} small dark matter
haloes form first, and then they subsequently merge  to form larger
ones.  At the same time,  gas cools and collapses  into the potential
well of these dark matter haloes  where star formation takes place,
giving rise to the first galaxies. 

Dwarf galaxies represent the low mass end of the cosmic assembly process,
and among them, satellite galaxies in the Local Group provide
the smallest-scale objects ($M_{\rm star} < 10^7 \Msun$) to test
our understanding of the galaxy formation process at the edge of its domain.

Small scales are a notoriously difficult test-bed for the most used
cosmological model, which is based on the cosmological constant
 \citep[$\Lambda$,][]{Riess1998,Perlmutter1999} that sets the current
 expansion of the Universe,  and the Cold Dark Matter paradigm \cite[CDM, e.g. ][]{Peebles1984}.
 Such a \LCDM model
has been challenged several times on the scales of dwarf galaxies, from the missing satellites
problem \cite[e.g.][]{Klypin1999,Moore1999}, to the cusp-core dichotomy  \cite[e.g.][]{Flores1994, Moore1994,
Oh2015} and more recently due to the so called "too-big-to-fail" problem \citep{Boylan-Kolchin2011} and the apparent 
planar configuration of satellite galaxies around the Andromeda galaxy and our own Milky Way \citep[][but see \cite{Buck2015,Buck2016}]{Ibata2013}.

All these problems arose from the (somewhat fallacious) comparison of
pure gravity (N-body) simulations with  real data based on the
observations of baryons. Thankfully in the recent years we have
witnessed a large improvement in our ability to simulate structure
formation including baryons both on large cosmological scales
\citep[e.g.][]{Schaye2015, Vogelsberger2014, Sawala2016, Sawala2016b} and for single objects
both on the scale of the Milky Way \citep[e.g.][]{Maccio2012, Stinson2013, Aumer2013, Hopkins2014, Marinacci2014, Wang2015, Dutton2015,
Wetzel2016}
and on the scale of dwarf galaxies \citep[e.g.][]{Governato2012, Simpson2013,Onorbe2015, Sawala2016b,Fitts2016}

These simulations have strongly increased our understanding of galaxy formation and alleviated, if not solved, most
of the \LCDM problems on small scales \citep[e.g.][]{Zolotov2012, Sawala2016, Tollet2016,Wetzel2016}

It is nowadays possible to run simulations of a single galaxy with
several million elements, and to do this  for several galaxies
covering a large fraction of their mass spectrum  \citep{Wang2015,Chan2015}.
Despite these advancements it is still  very
hard to attain such a resolution ($10^6$ elements) for the satellites
of our own Galaxy in a full cosmological context, since this will
require to resolve the whole object (Milky Way  + satellites) with
more than a billion particles. For comparison the best simulation
today of the Milky Way, the Latte project  \citep{Wetzel2016} has
achieved $\sim 10^7$ elements.

For this reason different approaches have been tried in the
literature.  In general the evolution of a satellite galaxy can be
split in two parts, its formation and evolution before being captured
by its final host (the isolation phase) and the accretion and
subsequent evolution within the host (the satellite phase).  In this
optic many authors have decided to somehow neglect the formation
process and to focus their attention  on the second phase, by studying
the effects of ram pressure and tidal effects of model (pre-cooked)
galaxies while orbiting their host \citep[e.g.][and references therein]{Kazantzidis2004,Mayer2006,Kang2008,Donghia2009,Chang2013, Kazantzidis2017}.

In this work we want to combine the two approaches, cosmological
simulations and simulations of galaxy accretion: namely we want to use
cosmological hydrodynamical simulations to generate the initial
conditions  for the isolated simulations of satellite galaxy
interaction. In this first paper (hereafter PaperI) we present the
results of a set of 27 high resolution simulations of galaxies forming
in dark matter haloes with masses between $5\times 10^8$ and $10^{10}
\Msun$. These simulations are meant to represent the properties of the
satellite galaxies {\it before} accretion, that we assume to happen at
$z=1$. In the second paper (Frings \etal 2017, hereafter PaperII) we
will study in detail the environmental effects (ram pressure, tidal
forces, mass removal, etc.) on the galaxy properties untill today's
time. 

The goal of our approach is to use more realistic initial conditions
to better understand the  effects of accretion and environment on the
evolution of the smallest galaxies we see today.

This paper is organized as follows, in section 2 we will introduce our
code and the hydrodynamical cosmological simulations used in this
paper. In section 3 we will present the evolution of our galaxies from
redshift $\sim 100$ to accretion redshift  that we set equal to one
\citep{Maccio2010b}.
We will concentrate on scaling relations between structural
parameters, on the dark matter, gas and stellar content and on the
relation between star formation and dark matter response.  In section
4 we will present our discussion and conclusion on this first phase of
the life of (future) satellites, while we leave to PaperII  the description of the accretion phase.

\section{Simulations}
\label{sec:setup}

\subsection{Initial Conditions}
\label{ssec:IC}

The simulations presented here are a series of fully cosmological
``zoom-in'' simulations of galaxy formation run in a flat $\Lambda$CDM
cosmology with parameters from the 7th year data release from the WMAP
satellite  \citep{Komatsu2011}: Hubble parameter $H_0$= 70.2 \kms
Mpc$^{-1}$, matter density $\Omegam=0.2748$, dark energy density
$\Omega_{\Lambda}=1-\Omegam -\Omegar=0.7252$, baryon density
$\Omegab=0.04572$, power spectrum normalization $\sigma_8 = 0.816$,
power spectrum slope $n=0.968$.  The haloes simulated in this paper
have been initially selected from two cosmological boxes of size
$L=10$ and $L=15$  \Mpch and initially run at two different resolutions either
with  $400^3$ with $350^3$ dark matter particles. 

We chose 27 haloes to be re-simulated at much higher
resolution using (depending on the mass) a zoom in factor of $8^3$ or $12^3$, and with the inclusion of
baryons.  We use a  modified version of the {\sc grafic2} package
\citep{Bertschinger2001} as described in \citet{Penzo2014} to create the
zoom-in initial conditions.
Depending on the initial number of particles (400$^3$ or 350$^3$), the box size
(10 or 15   \Mpch) and on the zoom level (8 or 12) we attain slightly different mass resolutions
for the different galaxies. In all cases we have about one million elements within the virial
radius at $z=1$.  The softening has been chosen to be $\approx 1/70$ of the intra-particle distance
\citep{Power2003} for the dark matter, and it has been rescaled for the gas as the square root of the mass difference 
in order to ensure a constant force resolution \citep{MosterMaccio2010}.
For the dark matter it ranges from 47 pc to 21 pc and from 21 to 9.4 pc for the gas. 
Our choice of softening is just a rescaling of the NIHAO resolution \citep{Wang2015}, which provides 
a convergence radius well below 1\% of the virial radius \citep[see discussion in][]{Tollet2016}.
Our high numerical resolution also ensures us that the half mass radius of the galaxy
is resolved with at least few hundred mass elements, allowing us to study the response
of dark matter to galaxy formation on the scales probed by the observations.
Table \ref{tab:res} contains all the needed information about the mass and space resolution. 

\begin{table}
\centering
\begin{tabular}{ccccc}
\hline
\hline
Res & m$_{\rm dm}$ & m$_{\rm gas}$ & $\epsilon_{\rm dm}$  & $\epsilon_{\rm gas}$ \\
\hline
  & M$_{\odot}$ & M$_{\odot}$ & pc & pc \\
\hline
1  &   $4.58 \times 10^3$   & $9.14 \times 10^2$  &  47 &  21 \\   
2  &   $2.02 \times 10^3$   & $4.04 \times 10^2$  &  31 &  14 \\   
3  &   $1.36 \times 10^3$   & $2.70 \times 10^2$  &  31 &  14 \\   
4  &   $6.00 \times 10^2$   & $1.19 \times 10^2$  &  21 &  9   \\   
\hline
\hline
\end{tabular}
\caption{The four different Mass and spatial resolution levels used in our simulations.}
\label{tab:res}
\end{table}

Star particles have an initial mass set to equal to $1/3$ of the gas particle mass,
They then return a fraction of their mass to the IGM via stellar winds
\citep[see][for more details]{Shen2010}, this implies that even our least
(stellar) massive galaxies are resolve with several hundred stellar particles.
Table \ref{tab:gal} contains the detailed information on the properties
of each galaxy, the last column refers to the corresponding resolution in table \ref{tab:res}.

\subsubsection{Central vs. Satellite-like initial conditions}

 Since we aim to compare our $z=1$ simulation results with the properties of local galactic satellites
one might wonder if selecting haloes that are actual satellites at $z=0$ (but still isolated at $z=1$) would 
make a difference w.r.t. using haloes that are isolated (centrals) also at $z=0$.

\begin{figure}
\includegraphics[width=85mm]{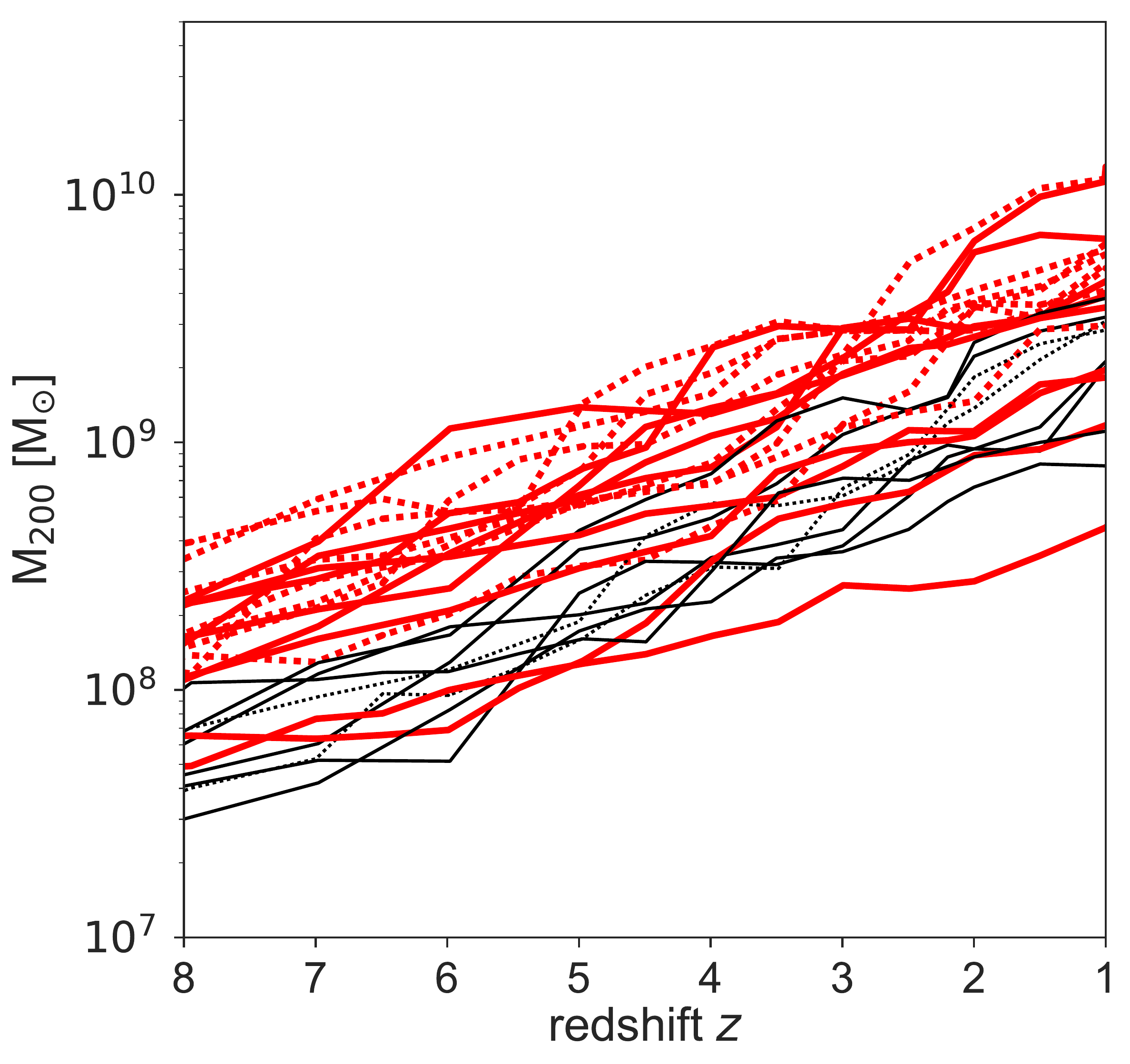}
\caption{The mass accretion history of our simulations. 
Solid (dotted) lines indicate haloes that are "centrals" ("satellites") at $z=0$.
Red is used for luminous satellites, while black represents dark ones.}
\label{fig:MAH}
\end{figure}

In order to test this point we have selected our haloes to be zoomed from two different
environments. Galaxies at resolution 2 and 4 have been selected to be isolated by $z=0$, meaning
they are the "central" object in their halo. Galaxies at resolution 1 and 3, on the contrary, are all
"satellites" of a more massive object (at least a factor or 20) by $z=0$. 
Despite the very different final environment, the two classes of objects have, 
at a fixed mass, very similar mass accretion histories  (up to $z=1$) and formation redshifts as shown 
in figure \ref{fig:MAH}.
As a consequence the luminous properties of the galaxies at  $z=1$ do not clearly separate in any of
the correlations we have studied in this paper; we have then decided to treat all haloes  as a single family.

\begin{table}
\centering
\begin{tabular}{lccccc}
\hline
\hline
M$_{\rm 200}$ & M$_{\rm star}$ & M$_{\rm gas}^{\rm cold}$  & $r_h^{2D}$ & $\sigma_v$ & Res \\
\hline
 M$_{\odot}$ & M$_{\odot}$ & M$_{\odot}$ & pc & km s$^{\rm -1}$ \\
\hline

1.13 $\times 10^{10}$ & 4.47 $\times 10^6$ & $8.62 \times 10^7$ & 337 & 12.3 & 3\\
1.03 $\times 10^{10}$ & 8.97 $\times 10^6$ & $2.44 \times 10^7$ & 630 & 15.1 & 3\\
6.33 $\times 10^{9}$ & 1.05 $\times 10^6$ & $1.73 \times 10^7$ & 148 & 9.3 & 1 \\
6.02 $\times 10^{9}$ & 3.74 $\times 10^6$ & $1.46 \times 10^7$ & 553 & 13.1 & 3 \\
5.72 $\times 10^{9}$ & 1.81 $\times 10^6$ & $3.22 \times 10^7$ & 243 & 8.3 & 3 \\
5.70 $\times 10^{9}$ & 1.19 $\times 10^6$ & $3.08 \times 10^5$ & 219 & 5.6 & 2 \\
5.55 $\times 10^{9}$ & 4.53 $\times 10^6$ & $1.31 \times 10^6$ & 394 & 10.6 & 2 \\
5.01 $\times 10^{9}$ & 4.59 $\times 10^4$ & $2.08 \times 10^6$ & 151 & 6.2 & 2 \\
4.48 $\times 10^{9}$ & 5.65 $\times 10^5$ & $5.48  \times 10^6$ & 301 & 10.6 & 3\\
4.45 $\times 10^{9}$ & 1.27 $\times 10^6$ & $1.29\times 10^7$ & 212 & 7.6 & 2 \\
4.09 $\times 10^{9}$ & 7.41 $\times 10^5$ & $9.78 \times 10^6$ & 172 & 8.6 & 2 \\
3.89 $\times 10^{9}$ & 1.29 $\times 10^6$ & $8.05\times 10^6$ & 260 & 11.3 & 2\\
3.81 $\times 10^{9}$ & 0  &  0  & -  & -  & 2 \\
3.61 $\times 10^{9}$ & 4.01 $\times 10^5$ & $5.86 \times 10^6$ & 152 & 9.45 & 2 \\
3.51 $\times 10^{9}$ & 4.07 $\times 10^5$ & $3.99 \times 10^5$ & 194 & 7.15 & 2 \\
3.19 $\times 10^{9}$ & 0  &  0  & -  & -  & 2 \\
3.08 $\times 10^{9}$ & 0  &  0  & -  & -  & 2 \\
2.95 $\times 10^{9}$ & 5.46 $\times 10^5$ & $3.33 \times 10^5$ & 172 & 8.9 & 3 \\
2.85 $\times 10^{9}$ & 0  &  0  & -  & -  & 3 \\
1.82 $\times 10^{9}$ & 1.45 $\times 10^5$ & $2.99 \times 10^3$ & 202 & 7.23 & 4 \\
1.70 $\times 10^{9}$ & 1.96 $\times 10^5$ & $2.43 \times 10^3$ & 77 & 3.4 & 4 \\
1.38 $\times 10^{9}$ & 0  &  6.1  $\times 10^4$& -  & -  & 4 \\
1.32 $\times 10^{9}$ & 0  &  4.81  $\times 10^2$& -  & -  & 4 \\
1.11 $\times 10^{9}$ & 0   & $1.18 \times 10^5$ & -  & -  & 2 \\
1.07 $\times 10^{9}$ & 4.48 $\times 10^4$ & $8.91 \times 10^5$ & 138 & 5.7 & 3 \\
8.01 $\times 10^{8}$ & 0  &  0  & -  & -  & 4 \\
4.50 $\times 10^{8}$ & 4.25 $\times 10^4$ & $4.14 \times 10^5$ & 102 & 4.8 & 3 \\
\hline
\hline
\end{tabular}
\caption{  Parameters of the simulated galaxies: total halo mass (M$_{\rm 200}$), stellar mass (M$_{\rm star}$), cold ($T<15000$) gas mass  (M$_{\rm gas}^{\rm cold}$), 2D half mass radius ($r_h^{2D}$), l.o.s. stellar velocity dispersion ($\sigma_v$). The last column (Res) indicates the resolution of the simulation, see table \ref{tab:res} for more details.  
Galaxies are listed in order of (decreasing) halo mass, dark haloes  have zero stellar mass.}
\label{tab:gal}
\end{table}

\subsection{Hydrodynamical code}
\label{ssec:Runs}

All haloes have been run only to $z=1$, since after this time we
assume that the haloes will be accreted onto a more massive object,
becoming hence satellites \citep[e.g.][]{Maccio2010b}.
The evolution of our haloes after accretion
is described in PaperII. For this paper we will
only present the properties of our simulated galaxies at $z=1$ or
earlier times. 

The simulations have been performed with the SPH code
{\sc gasoline} \citep{Wadsley2004}. The setup of the code is the same
as the one used in the MaGICC papers \citep{Maccio2012, Stinson2013,
  Kannan2014}, the code includes metal cooling, chemical enrichment,
star formation, feedback from massive stars and super novae (SN).
Stars are formed from gas cooler than $T$ = 15000 K, and denser than
$n_{\rm th}=60 {\rm cm}^{-3}$, this number represents the density of a
``kernel'' of particles (32) inside a sphere of radius equal to the
softening \citep[see][]{Wang2015}. We adopt a star
formation efficiency parameter $c_\star$=0.1.   The cooling used in
this paper is described in detail in \citet{Shen2010} and includes
photoionization and heating from the \citet{Haardt2012}
ultraviolet (UV) background, Compton cooling, and Hydrogen, Helium and
metal line cooling. 

SN feedback is implemented using the blastwave approach as
described  in \citet{Stinson2013}, which relies on delaying the cooling
for particles near the SN event. We also add what we dubbed Early
Stellar Feedback, namely we  inject as thermal energy $10\%$ the UV
luminosity of the stars before any SN events take place,  without
disabling the cooling \citep[see ][for more details]{Stinson2013}. 

Haloes in our zoom-in simulations were identified using the MPI+OpenMP
hybrid halo finder {\sc ahf}\footnote{http://popia.ft.uam.es/AMIGA}
\citet{ahf}.   The virial masses of the
haloes are defined as the mass within a sphere containing
$\Delta=200$ times the cosmic critical matter density. The virial
(total) mass is denoted as $M_{200}$, the virial radius as $r_{200}$,
finally $M_{\rm star}$ indicates the total stellar mass within
0.1$r_{200}$.

\section{Results}
\label{sec:results}

In the following we will present the results of our simulated
galaxies.  In all plots simulations are always shown at $z=1$, which
represents the time at which (on average) these haloes will be
accreted onto a more massive  halo. In most of the plots the
simulation results are represented by color dots or lines. The same
color corresponds to the same galaxy in all plots, making easier to
connect the  different properties of the same galaxy across the
various figures.

\begin{figure}
\includegraphics[width=85mm]{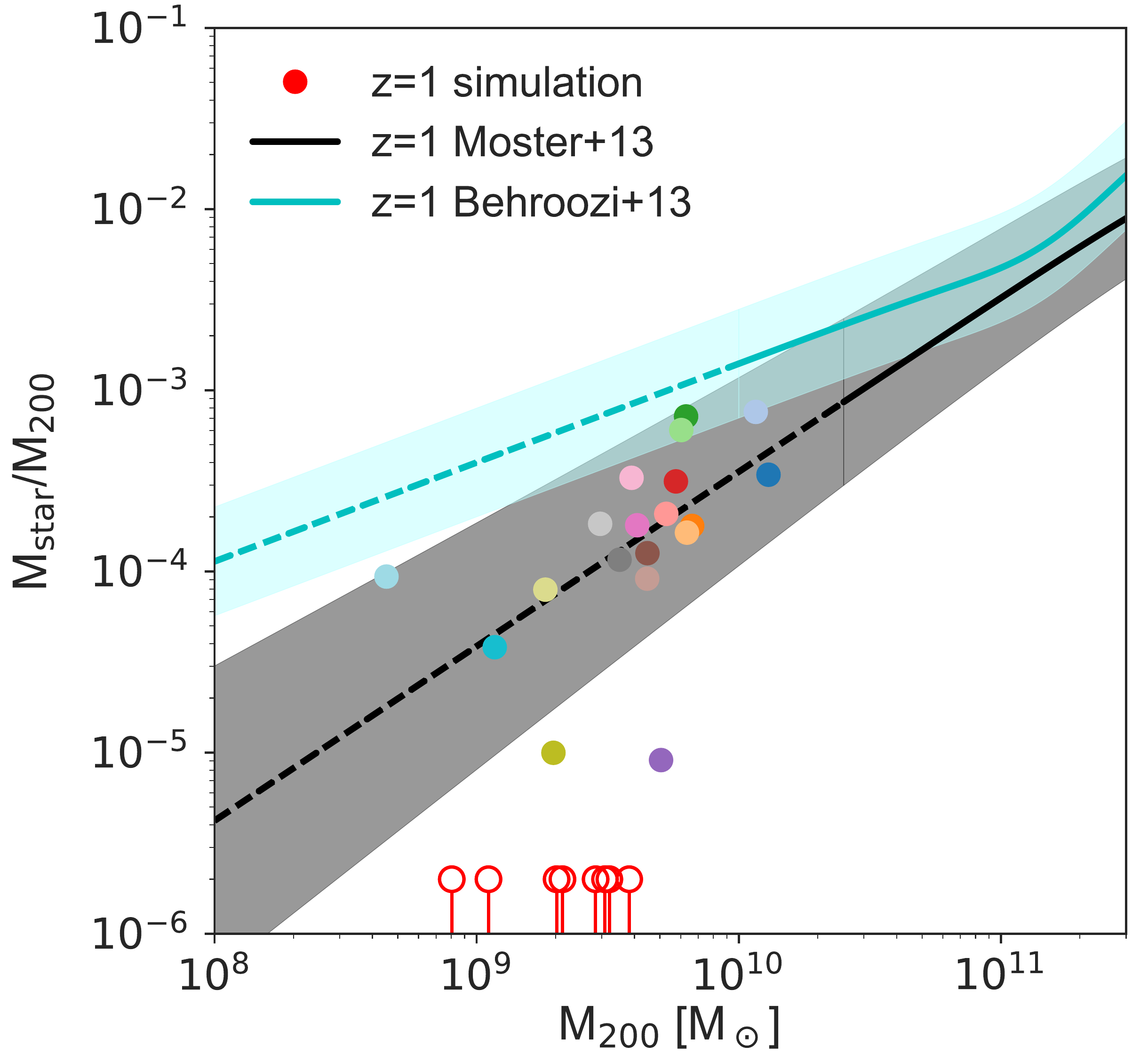}
\caption{The stellar mass halo mass relation for our simulated
  galaxies at $z=1$. The colorful points are haloes with stars, while
  the red empty circles represent dark haloes. The abundance matching
  relations from \citet{Moster2013} and \citet{Behroozi2013} are
  shown in black and cyan respectively, the dashed lines indicate the
  extrapolation to lower masses.}
\label{fig:AM}
\end{figure}

\subsection{Dark, stellar and gas masses at z=1}

At first we look into the relation between stellar mass and halo mass
for our galaxies. Results are shown in figure \ref{fig:AM}, where
colorful circles represent haloes that did form stars, while empty red
circles show ``dark'' haloes.  The grey and cyan lines represent the
abundance matching relations from \citet{Moster2013} and \citet{Behroozi2013},
respectively. 
All our galaxies seem to prefer a lower stellar mass than what
is predicted by Behroozi and  collaborators and being more in agreement
with a simple extrapolation of the Moster relation (the extrapolated
part is marked by a dashed line in both cases).

An interesting thing to notice is the very large scatter (0.45 dex)
in stellar mass at a fixed halo mass: for example for a halo mass around
$7\times 10^9$ \Msun the ratio between stellar mass and halo mass
changes by about two  orders of magnitude from $10^{-3}$ to $10^{-5}$.
We will return to the origin of this scatter later in section
\ref{sec:SFH}.

For halo masses below  $4 \times 10^9 \Msun$ about half of the haloes
remain dark,  in other words they are not able to create a single
stellar particle.  This is in fairly good agreement with
previous results of \citet [][see also \cite{Simpson2013}]{Sawala2016b} which use  several hydrodynamic cosmological simulations of the Local Group to 
study the discriminating factors for galaxy formation (i.e. being luminous) in low mass haloes.
Based on their (larger) sample of haloes they also found about half 
of the haloes remaining dark at $z=1$ at these mass scales.
Such a dark fraction is also consistent  with what is required
to solve the so-called missing satellite problem
\citep{Klypin1999,Maccio2010, Sawala2016b}.   Despite the large number of dark
haloes, it is interesting to notice that the lowest mass  halo in our
sample (cyan point with $M_{200} = 5 \times 10^8 \Msun$) is nevertheless
luminous with about $10^4 \Msun$ of stars.

\begin{figure}
\includegraphics[width=85mm]{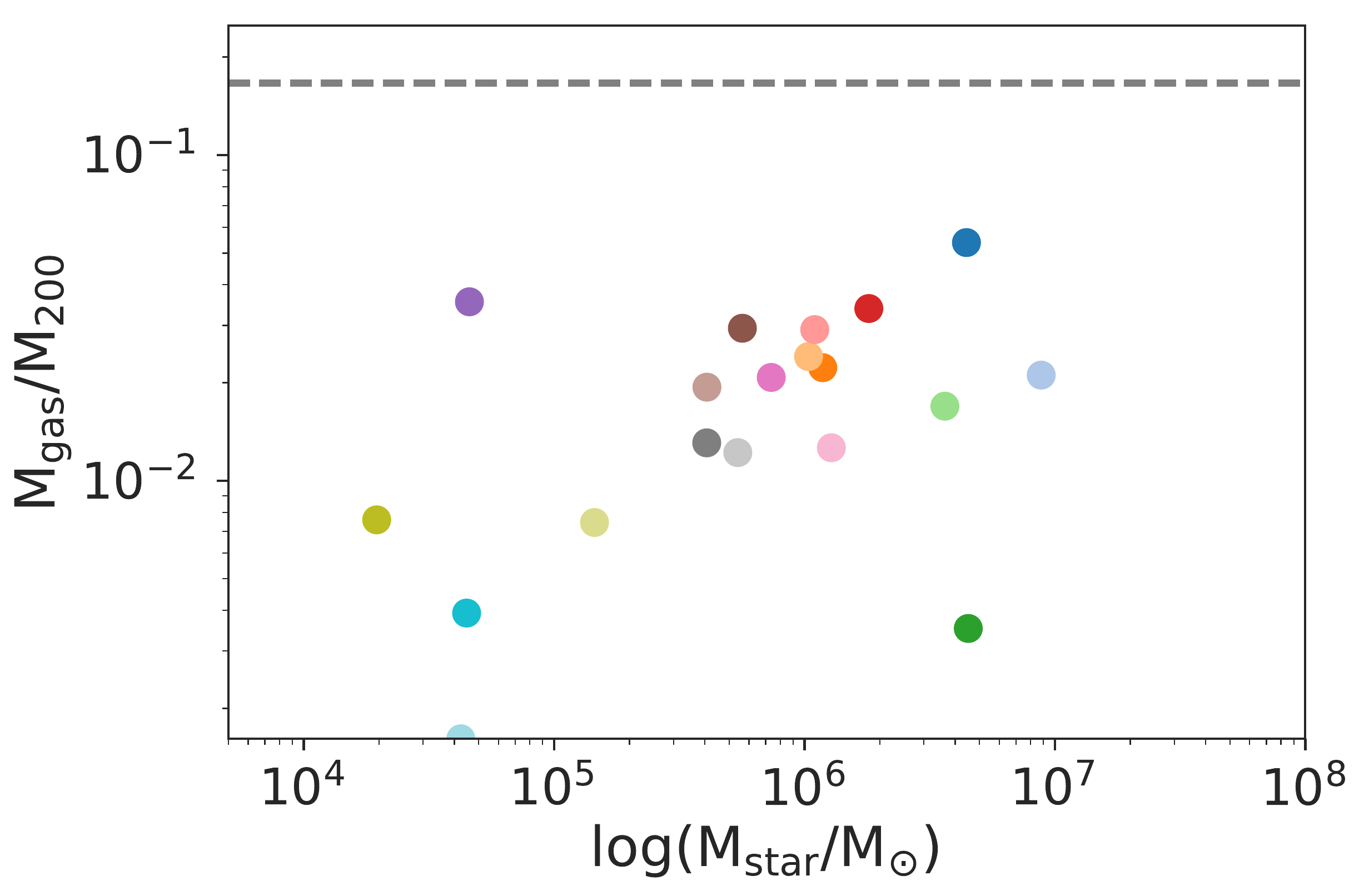}
\caption{The gas to total mass ratio as a function of stellar
  mass. The grey dashed line represents the cosmic value
  $\Omega_b/\Omega_m=0.155$ for the WMAP7 cosmology. The galaxy color
  coding is the same as in figure \ref{fig:AM}.}
\label{fig:gasfrac}
\end{figure}

At the mass scales analyzed in this paper, we expect galaxies to be
quite inefficient in accreting baryons due to the  UV
background \citep[e.g.][]{Gnedin2000,Hoeft2006,Okamoto2008, Simpson2013,Noh2014}. In figure \ref{fig:gasfrac} we
show  the gas to total mass fraction as a function of stellar
mass. All galaxies are strongly baryon ``deprived'' with  respect to
the universal baryon fraction (represented by the grey dashed line),
with some galaxies able to  accrete (and retain) less than 10\% of the
available baryonic budget, almost regardless of their stellar mass.

On the other hand the very low gas fraction could also be a result of
gas outflows due to SNe, given that the mass loading factor of
winds increases at lower circular velocities \citep{Dutton2012}.

Such question has been raised before, for example \cite{Simpson2013} used
a set of AMR (adaptive mesh refinement) cosmological simulations to study the effect
of reionization on the gas fraction of haloes with masses about $10^9 \Msun$.
They found that reionization  is primarily responsible for 
preventing gas accretion in their simulations.

\begin{figure}
\includegraphics[width=85mm]{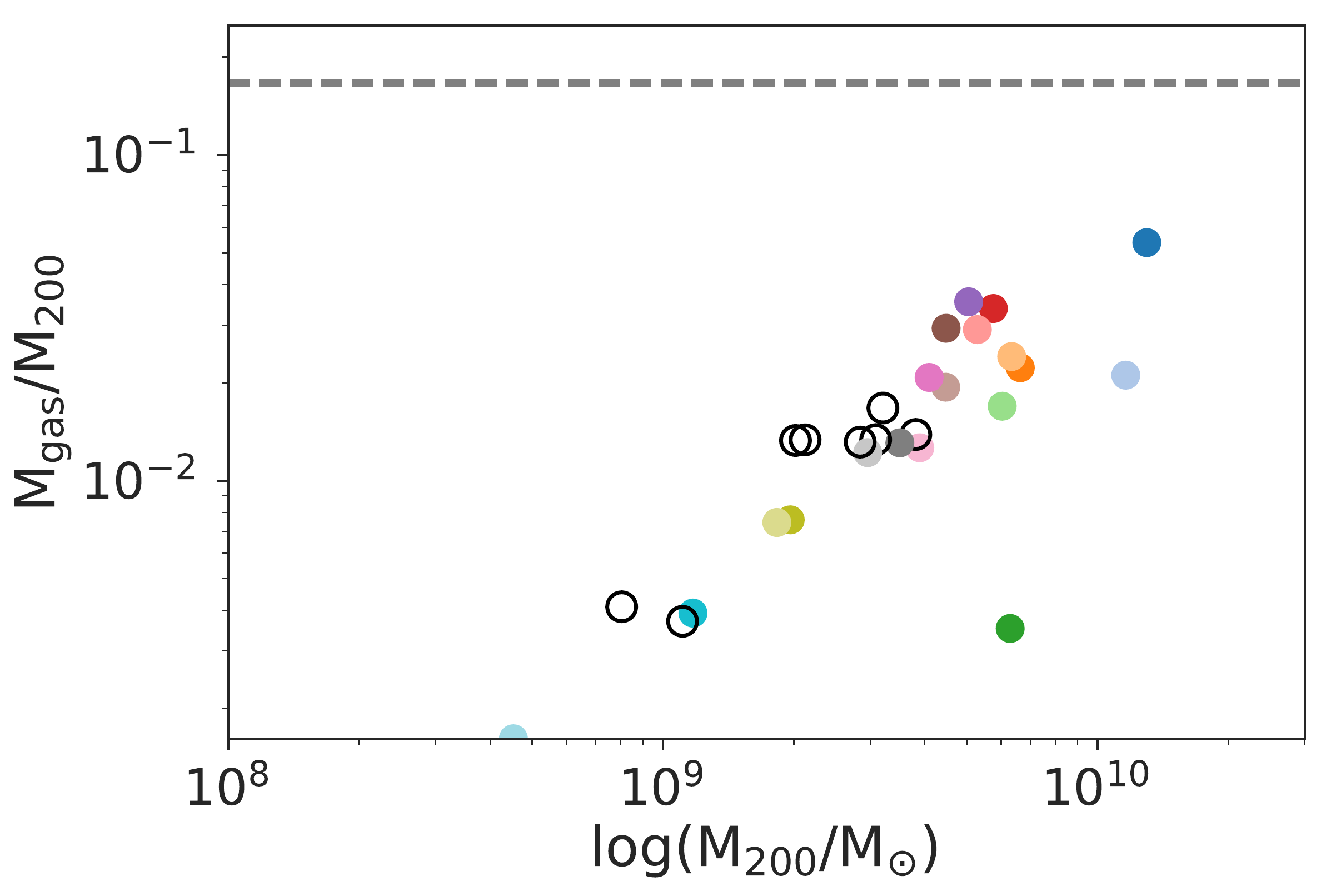}
\caption{The gas to total mass ratio as a function of virial
  mass. The grey dashed line represents the cosmic value
  $\Omega_b/\Omega_m=0.155$ for the WMAP7
  cosmology. The galaxy color coding is the same as in figure
  \ref{fig:AM}, while empty circles represent dark haloes (i.e. haloes
  that did not form stars).}
\label{fig:gasfracall}
\end{figure}

In our case we  can use the ``dark'' haloes, i.e. haloes that did not form any stars, to
also address this question, since obviously they have been affected by the UV
background but not by SN explosions.  In figure \ref{fig:gasfracall}
we plot the gas fraction as a function of the virial mass of the
halo.  Dark haloes (represented by empty symbols) have similar gas
fractions as luminous haloes with the same total mass, strongly
suggesting that the UV background is the main reason for the lack  of
baryons at these mass scales in good agreement with previous studies
\citep{Simpson2013, Sawala2016b}

\begin{figure}
\includegraphics[width=85mm]{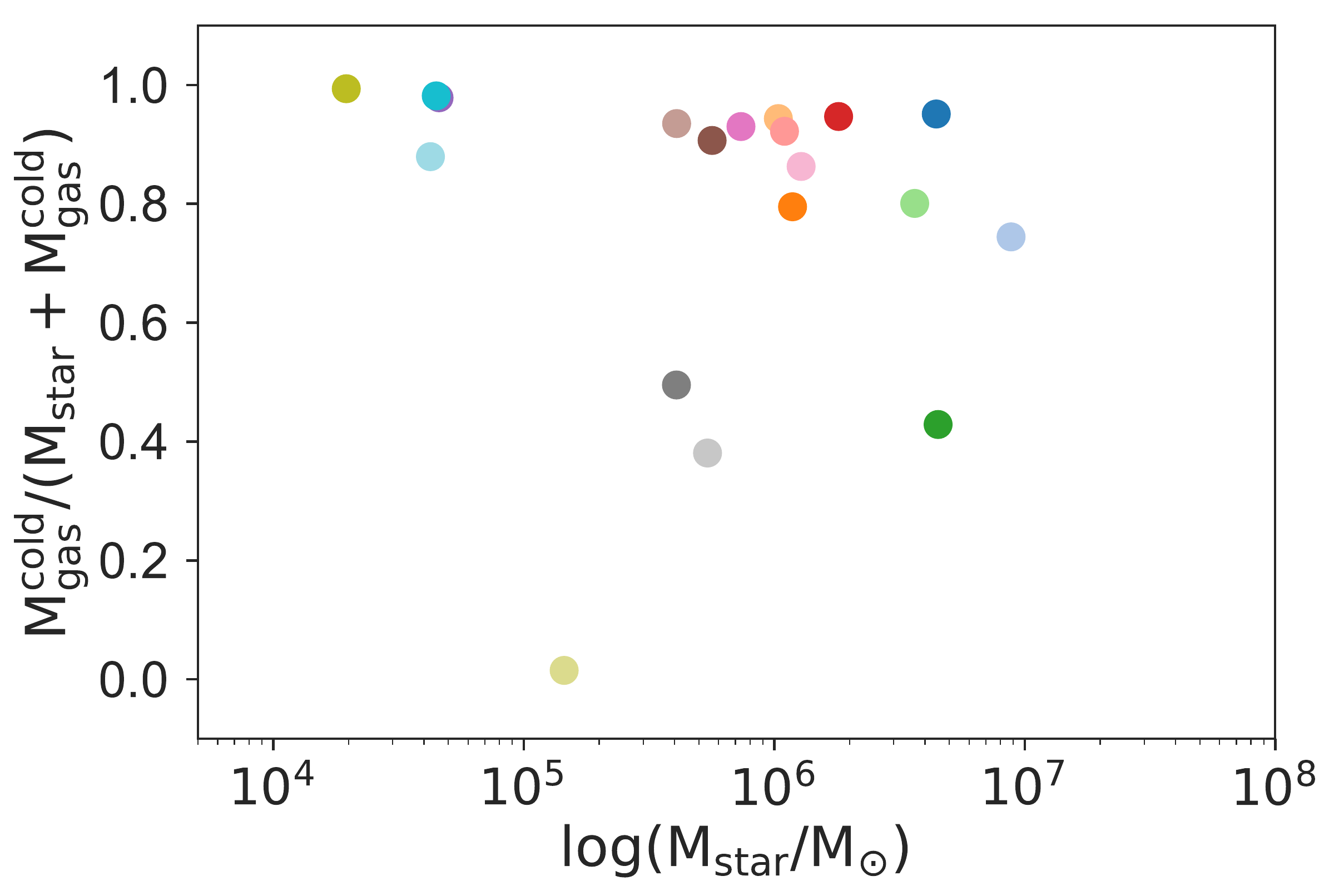}
\caption{The cold gas mass fraction as function of stellar mass. Cold
  gas is defined as gas with $T<15000$K.  The galaxy color coding is
  the same as in figure \ref{fig:AM}.}
\label{fig:coldgas}
\end{figure}

Our galaxies are quite inefficient in converting their (cold) gas
into stars, as shown in figure \ref{fig:coldgas}, where we plot the
cold gas fraction, defined as the mass in gas with $T<15000$ and
hence eligible  for star formation, as a function of the total stellar
mass. Most galaxies have four to six times more cold gas than stars at
$z=1$. Since today's Milky Way and M31 satellites are very gas poor,
this implies that environmental transformation (e.g. ram pressure
stripping) should play an important role in removing gas and
completely quenching these galaxies after accretion (see PaperII).

\subsection{Galaxy properties and scaling relations}

Despite that our simulations are for isolated haloes, their aim is to
predict the properties of the progenitor (pre-infall)  of galactic
satellites.  It then makes sense to compare their structural
parameters (radius, velocity dispersion etc.)  at $z=1$, before
infall, with the observations of Milky-Way and Andromeda satellites.

For the observational data we have used a compilation from
M. Collins (private communication) which includes results
for the Milky-Way and the Andromeda (M31) galaxy satellites including:
\citet[][MW]{Walker2009}, \citet[][MW]{Koposov2011}, \citet[][M31]{Tollerud2012,Tollerud2013}, \citet[][M31]{Ho2012},
\citet[][M31]{Collins2013}, \citet[][M31]{Collins2013}, \citet[][MW]{Kirby2013},
and \citet[][M31]{Martin2014}.

\begin{figure}
\includegraphics[width=85mm]{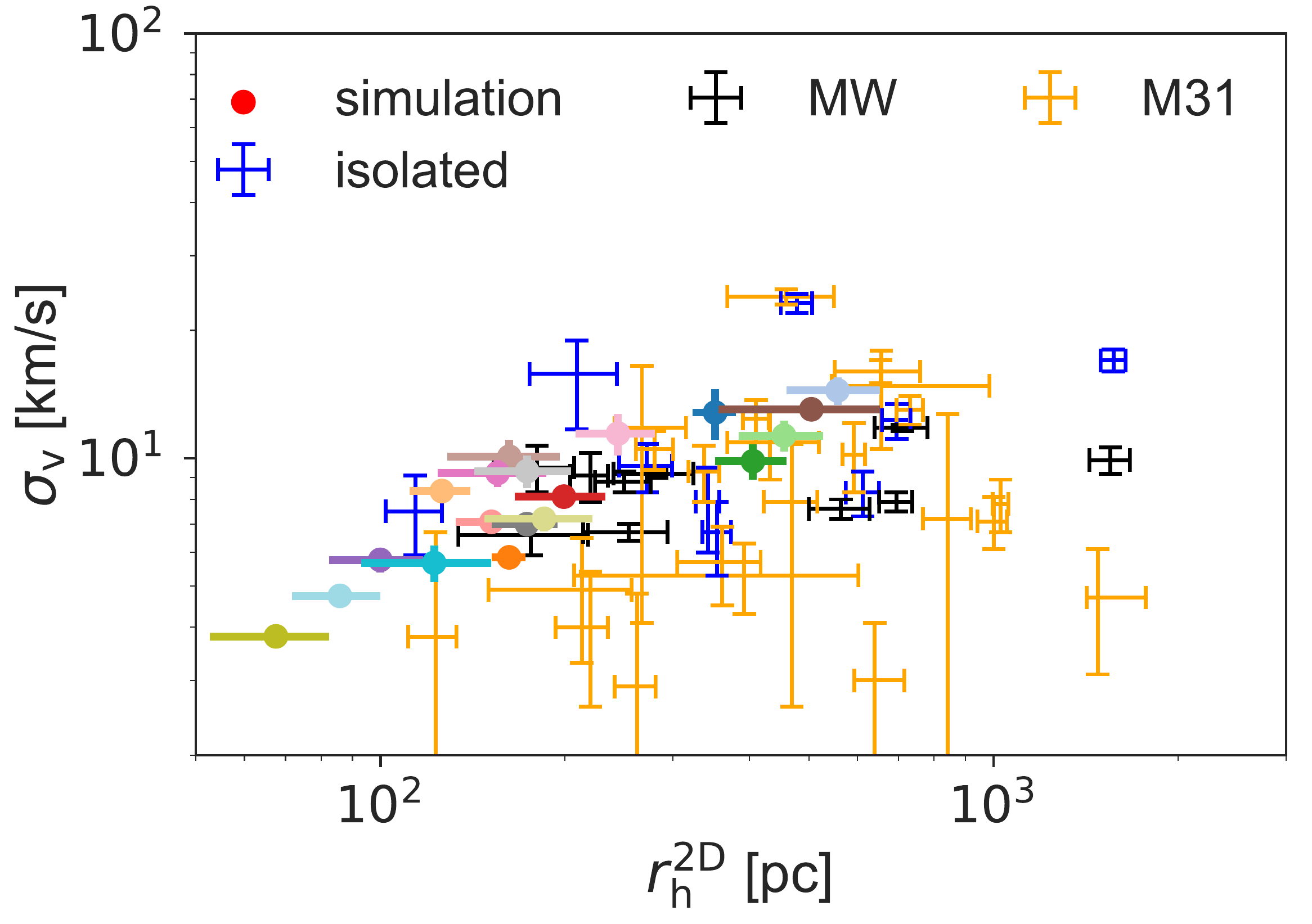}
\caption{ The 2D half mass radius vs. the line of sight stellar velocity dispersion. Simulation results are color coded as in figure  \ref{fig:AM}, the error bars
represent the $1-\sigma$ scatter from ten different projections. Crosses with error bars represent  observational data from \citet{Collins2013}, see text for more details.}
\label{fig:rhalf}
\end{figure}

In figure \ref{fig:rhalf}  we show the relation between the half
(stellar) mass radius and the stellar  velocity dispersion. In order
to mimic observations, the $r_{\rm h}^{2D}$ has been computed in two
dimensions,  meaning that we randomly project each galaxy and then
compute the half mass radius using 2D shells; the velocity dispersion
is computed along the projection radius of the satellite, equivalent to a line
of sight velocity dispersion.  We then repeat this procedure ten times for each galaxy and
show the average result  and its one sigma scatter for every galaxy.

The simulations reproduce quite nicely the trend of
the observations especially for isolated haloes
(blue crosses), but they have a smaller scatter than the one observed in
the MW and M31.  We will show in PaperII, that this scatter increases
substantially after the satellite is  accreted and tidally perturbed. 

\begin{figure}
\includegraphics[width=85mm]{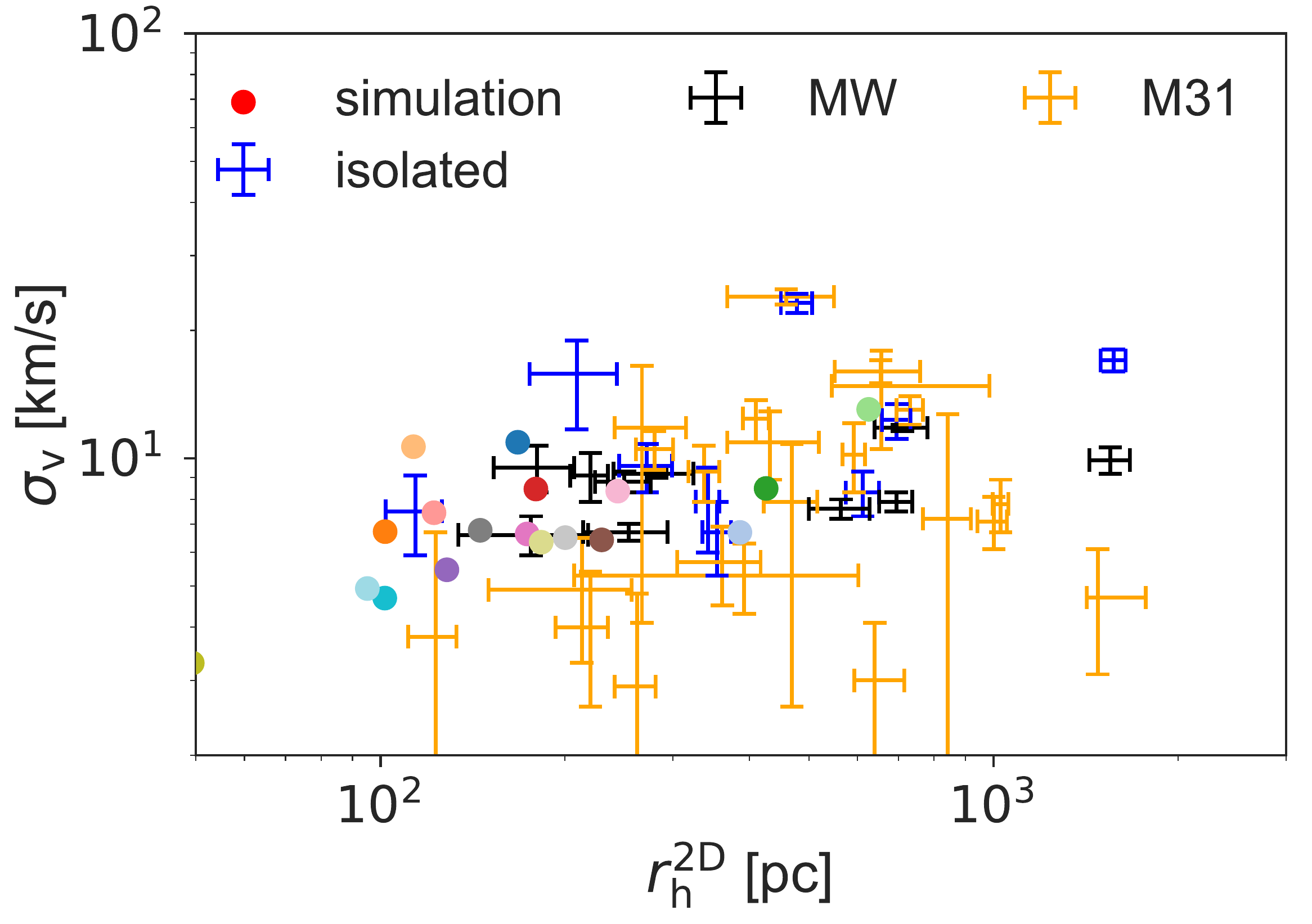}
\caption{ Same as figure \ref{fig:rhalf} but using a random redshift between 1 and 3 to compute the properties of the galaxies.}
\label{fig:rhalf_random}
\end{figure}

In figure  \ref{fig:rhalf}  the properties of the galaxies are shown at $z=1$ which we assume to be the accretion redshift
for all satellites. This is clearly a quite strong assumption since both observations and simulations do show 
a quite large scatter in the satellite accretion redshift \citep{Maccio2010b,Weisz2014}.
In order to test the impact of our {\it single accretion redshift} approach, in figure \ref{fig:rhalf_random} we show the 
same quantities as in figure  \ref{fig:rhalf}, but this time we have computed them at a different redshift (chosen random between
1 and 3) for every galaxy. 
While some points do move, we do not see any particular  systematic change in the plot, which
makes us confident that we are not introducing any strong bias by performing our  study at $z=1$.

\begin{figure}
\includegraphics[width=85mm]{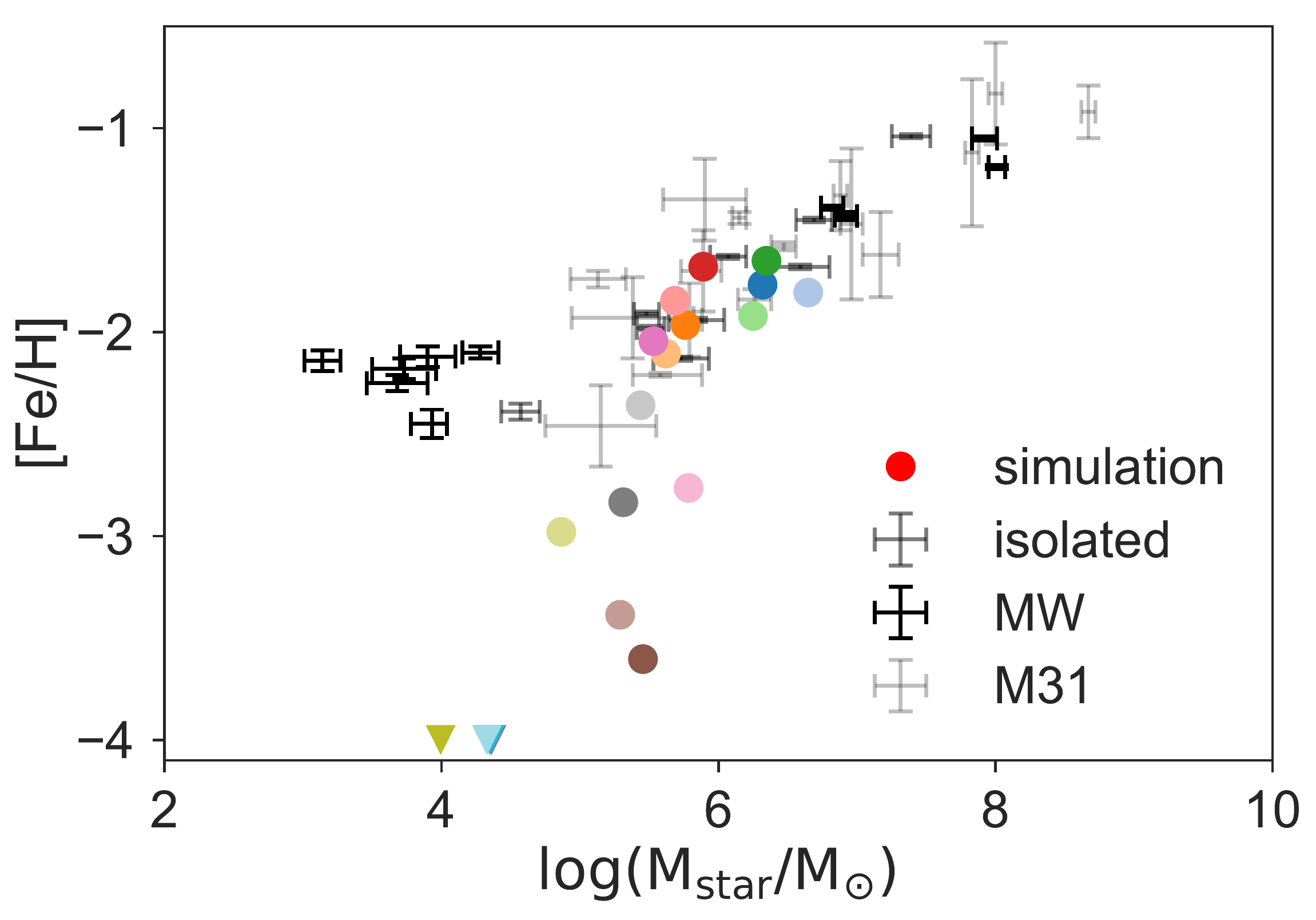}
\caption{ Stellar mass - metallicity relation. Observations from
  \citet{Kirby2011,Kirby2013} are represented by  black and grey symbols with
  error-bars. Simulation results are color coded as in figure  \ref{fig:AM}; the triangles represent an upper limit
  to the satellite metallicity.}
\label{fig:metals}
\end{figure}

Figure \ref{fig:metals} shows the relation between stellar mass and
metallicity.   The simulated galaxies have the same color coding as in
figure \ref{fig:AM} and are compared with observations from  \citet{Kirby2011,Kirby2013}.
Down to a stellar mass  of about $10^6$ \Msun there is a
fairly good agreement between  simulations and observations. 
Between
stellar masses of $10^5$ \Msun and $10^6$ \Msun  
the simulations start to have  lower metallicity  compared
with observations, and below $10^5$ \Msun  they have too 
low metallicities, by more than two dex. 
We ascribed this difference to the inability of our
enrichment algorithms to cope with very rapid and very small star
formation bursts.  At these very low stellar masses, our galaxies have
a very rapid single stellar burst (see  figure \ref{fig:sfr}) which
happens on time scales comparable to our internal time stepping.  This
means that the time resolution is too short to properly enrich the gas
and hence the very low metallicity. We plan to improve our chemistry
network and revisit this issue in a future publication. 

The ability of our more luminous galaxies to match the stellar mass
metallicity relation suggests that this  relation is already in place
{\it before} the infall and that tidal effects will make the galaxies
move along the relation, see PaperII for more details.

\subsection{Star formation rate and halo response}

\begin{figure*}
\includegraphics[width=160mm]{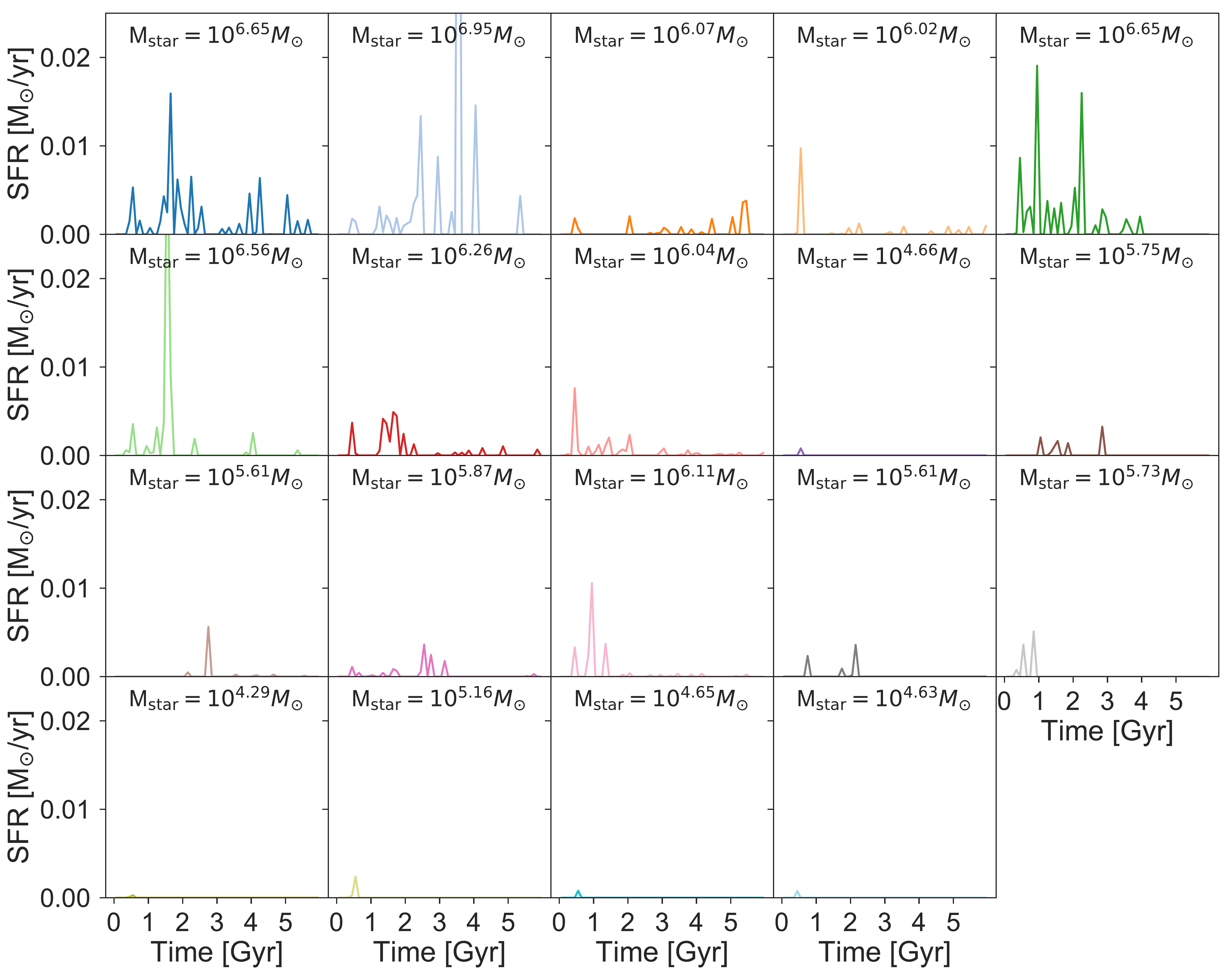}
\caption{Galaxy star formation histories computed over a period of 100 Myrs. 
The galaxies are ordered with decreasing total mass (as in table \ref{tab:gal}). 
The values of the stellar masses are reported in each single box. The color coding is the same as in figure \ref{fig:AM}.}
\label{fig:sfr}
\end{figure*}

As already mentioned above the star formation rate at the edge of
galaxy formation  is quite stochastic and made of rapid bursts,
followed by long quiescent periods,  as shown in  figure
\ref{fig:sfr}, where the star formation is computed over a period of 100 Myrs. 
In this figure galaxies are ordered by halo mass (according to table \ref{tab:gal}) but retain the
same coloring scheme as in figure  \ref{fig:AM}.

\begin{figure}
\includegraphics[width=85mm]{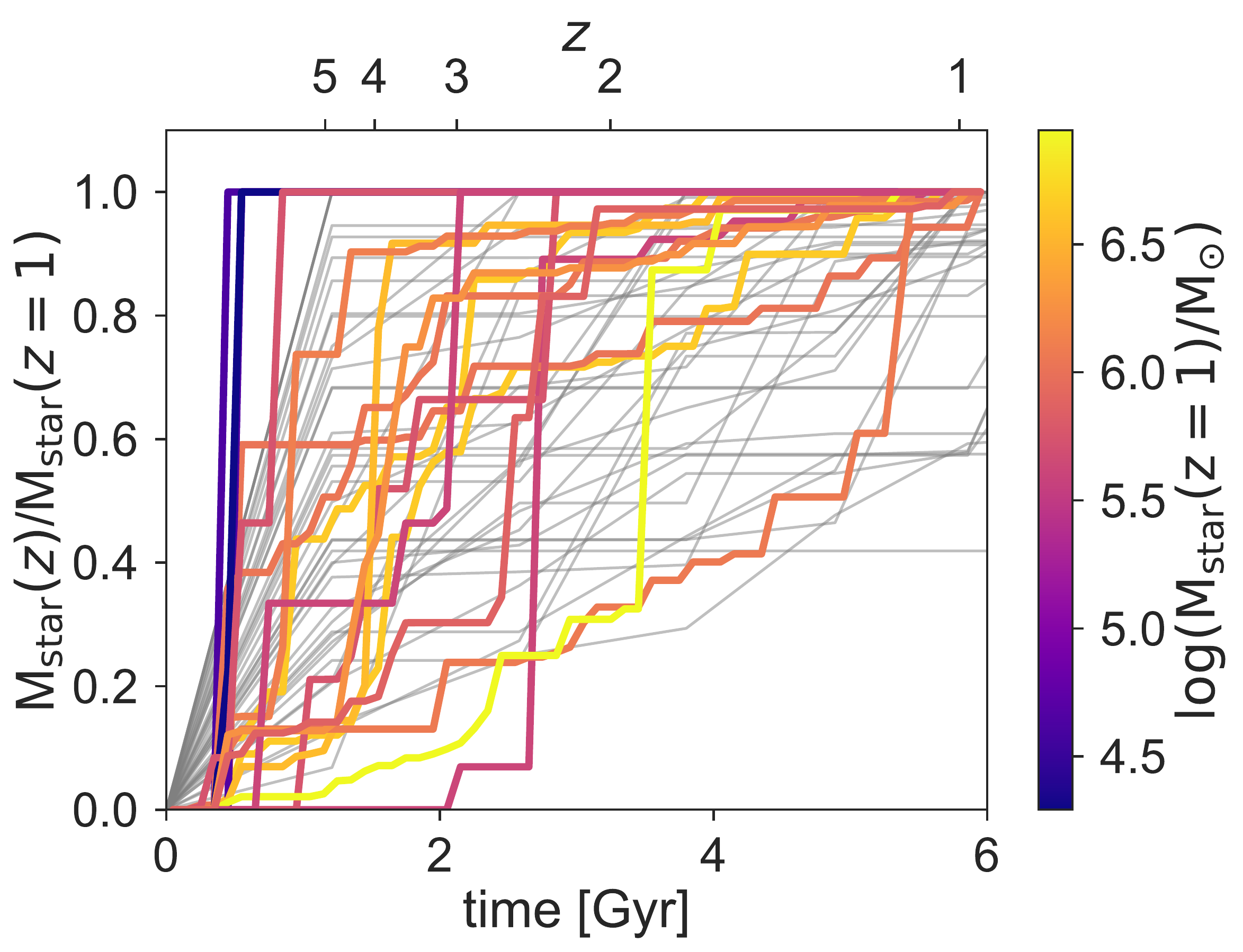}
\caption{Cumulative stellar mass growth up to $z=1$. Simulations are color coded according to their final stellar
  mass (right side bar). Observations from \citet{Weisz2014} are also limited to $z=1$ and are shown in grey.}
\label{fig:sfr2}
\end{figure}

Galaxies with similar halo mass (i.e. in neighboring panels in figure \ref{fig:sfr})
show quite diverse star formation
histories, with  different times and intensities for the  stellar
bursts.  This is in quite good agreement with the diversity in the stellar
mass assembly of satellites in the Milky Way and M31, as observed by
\citet{Weisz2014}. In figure \ref{fig:sfr2} we directly compare  the
cumulative stellar mass growth of our simulated galaxies (color coded
according to their stellar mass) with the results of Weisz and
collaborators up to $z=1$. As already noted in previous simulations
\citep{Governato2015,Fitts2016,Wetzel2016} we are also able to nicely reproduce
the diversity of the observed dwarf galaxies star formation histories.

\begin{figure}
\includegraphics[width=85mm]{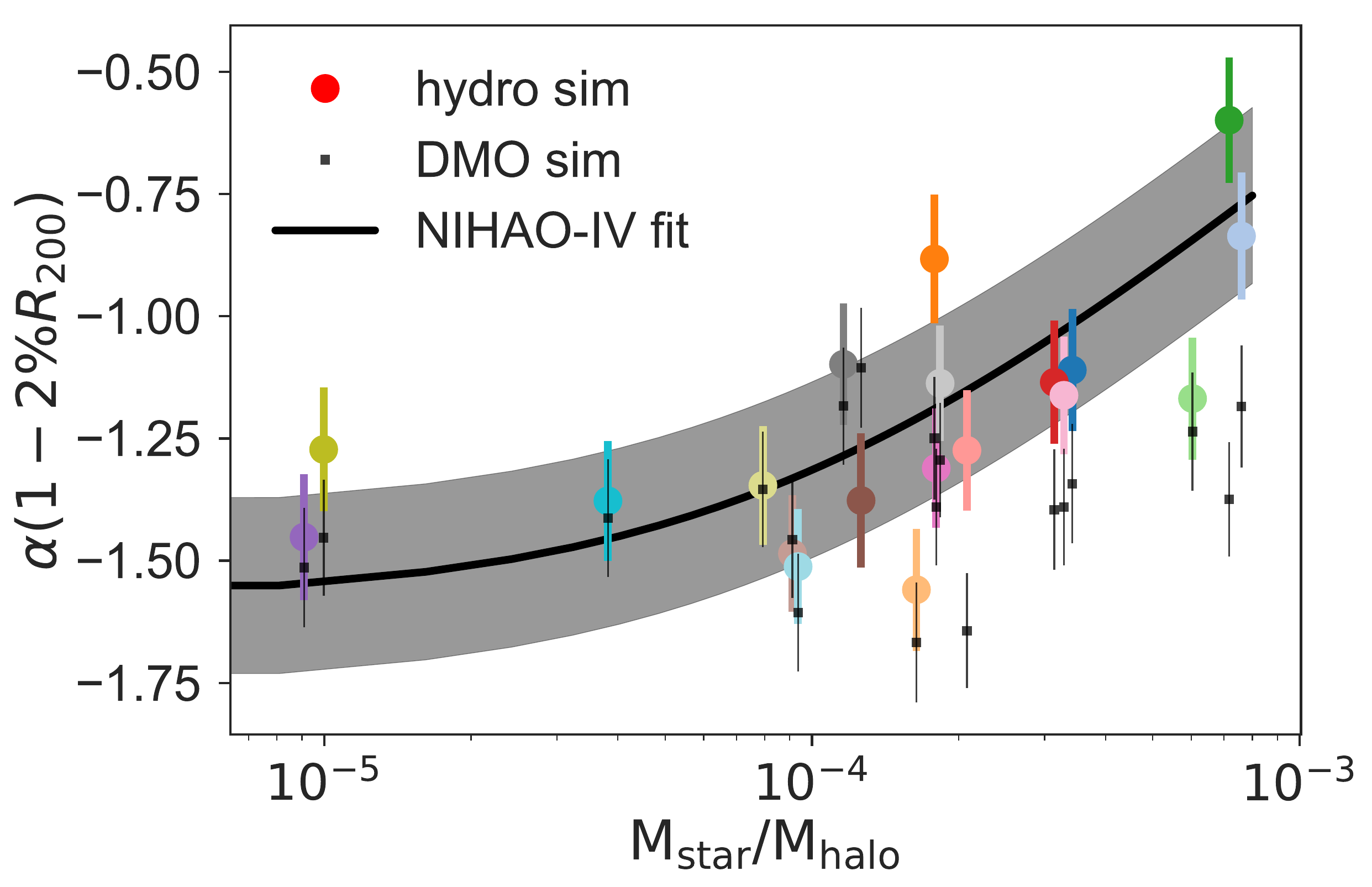}
\caption{Logarithmic slope of the dark matter halo profiles in the
  hydro simulation (color symbols) and in the dark matter only
  simulation (black dots).  The solid line shows the fitting formula
  proposed in the NIHAO-IV paper (Tollet \etal 2016). The slope
  $\alpha$ is computed between 1 and 2\% of the virial radius, while the error bars represent the
uncertainty from the fitting routine.}
\label{fig:alpha}
\end{figure}

Several recent papers have pointed out a correlation between  repeated
gas outflows due to star formation  bursts and the expansion of the
inner dark matter distribution 
\citep{Pontzen2012,Maccio2012, DiCintio2014a, Madau2014, Chan2015, Dutton2016b,Read2016, Tollet2016}.
It is then interesting to look at the inner slope of the dark matter density profile
in the hydro simulations and to compare this to their Dark Matter Only
(DMO) counterparts.  In figure \ref{fig:alpha} we show the logarithmic
slope of the dark matter density profile  ($\alpha = d \log (\rho) / d
\log r$) computed between 1 and 2\% of the virial radius, versus the
stellar mass to halo mass ratio.  We chose the latter quantity since
it has been shown to be the most correlated with $\alpha$ \citep{DiCintio2014a}.
Results from hydro simulations are represented
by the usual color symbols while the corresponding DMO (i.e. Nbody) results are
shown as black squares; in both cases the error bars represent the
uncertainty from the fitting routine.  In the same plot we show the
fitting formula from \citet{Tollet2016}, which was based on the
analysis of 90 galaxies from the NIHAO suite \citep{Wang2015}.

There is a quite good agreement with the results from \citet{Tollet2016},
\citep[see also][]{Chan2015}, meaning that we see a partial halo
expansion for a star formation efficiency
($M_{\rm star}/M_{\rm  200}$) close to $10^{-3}$ (the first two points on the right), but
then for lower star formation efficiency our galaxies retain the same
dark matter density profiles as their pure N-body counterparts. Baryons
are able to alter dark matter profiles possibly only  in very massive
satellites, while smaller objects are supposed to retain the typical
CDM cuspy Einasto-like profiles  \citep{Dutton2014} as already pointed
out in several previous studies like \cite[e.g.][and references therein]{Governato2012, Onorbe2015}

\begin{figure}
\includegraphics[width=85mm]{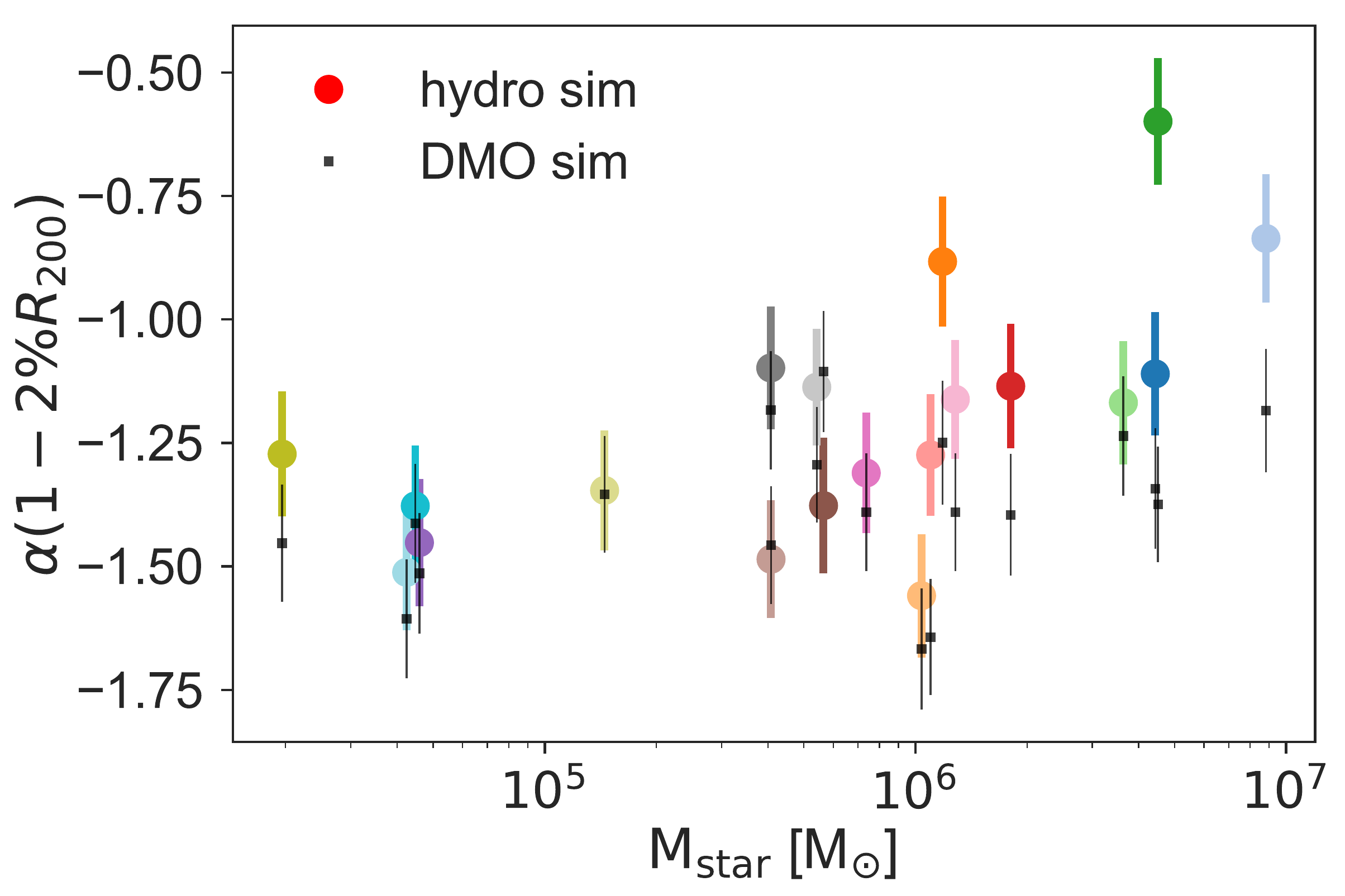}
\caption{Logarithmic slope of the dark matter halo profiles in the
  hydro simulation (color symbols) and in the dark matter only
  simulation (black dots) as  a function of galaxy stellar mass.  The
  slope $\alpha$ is computed between 1 and 2\% of the virial radius.}
\label{fig:alpha2}
\end{figure}

In figure \ref{fig:alpha2} following \cite{Governato2012} we also show the same slope $\alpha$ as a function of stellar
mass to facilitate a possible comparison with observations, symbols
have the same meaning as in figure \ref{fig:alpha}.

\begin{figure}
\includegraphics[width=85mm]{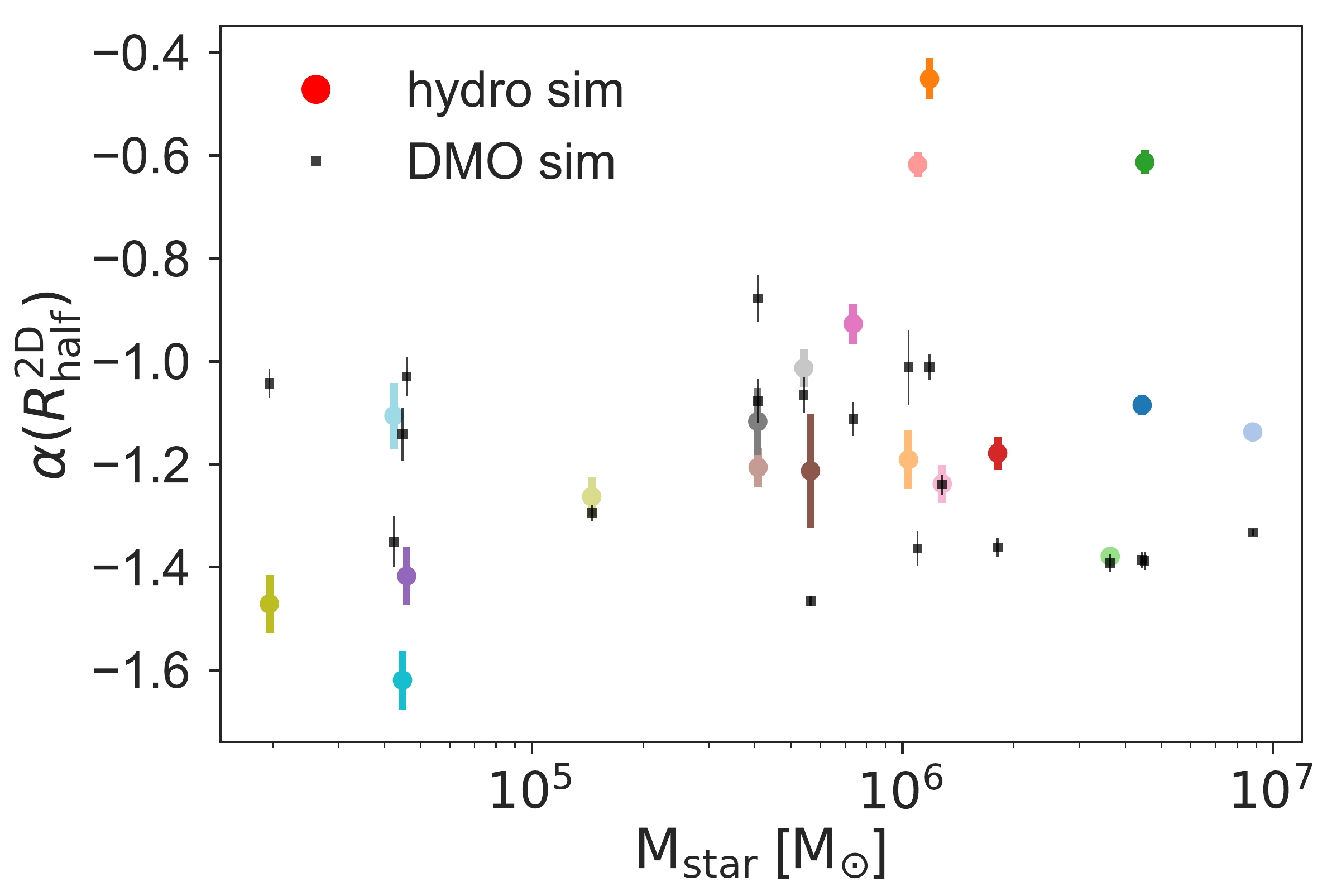}
\caption{Logarithmic slope of the dark matter halo profiles at the scale of the 
half mass radius. Color symbols represent the hydro simulation (same scheme as previous plots), the black  squares the dark matter only runs.}
\label{fig:alpha3}
\end{figure}

Our choice of measuring the profile slope between 1 and 2\% of the virial radius is somehow
arbitrary; for this reason we also compute it at a more natural length scale as the 2D half mass radius ($r_h^{2D}$,
for this measurement we used five equally spaced logarithmic bins around the radius.).
Figure \ref{fig:alpha3} shows the behavior of this new measurement of alpha as a function of the galaxy stellar
mass. 
By comparing the slope for the Dark Matter Only (DMO) simulations with the one of the Hydro ones, also in this case there seems 
to be a particular stellar mass (around $M_{\rm star} \approx 10^6 \Msun$) above which the dark matter profiles
becomes flatter in the hydro simulations. It is also interesting to note that for very low stellar masses, the 
hydro simulations are slightly contracted w.r.t. N-body ones on the scale of $r_h^{2D}$.

Recently \cite{Read2016} presented very high resolution simulations of isolated dwarf galaxies
reporting that if star formation proceeds for long enough,  dark cores of size
comparable to $r_h^{2D}$ always form. 
The key factor is to have an extended star formation period, of about 4 Gyr for a $10^8$ \Msun halo 
and $14$ Gyr for $10^9$ \Msun one.
Our galaxies from one side seem to support Read \etal findings in a sense that galaxies with "continuous"
star formation do seem to have flat profiles at the half mass radius, as can be seen by looking at the galaxies in the first
row of figure \ref{fig:sfr} and their respective position in figure \ref{fig:alpha3}.
On the other hand none of our low (stellar) mass galaxies has a cored profile, not even at the half mass radius.
This could be an indication that in a more realistic, cosmological set up (which also includes the UV background, an ingredient
missing in Read \etal) all star formation histories are indeed truncated after the first bursts
and no cores should be expected in low mass galaxies. 

A corollary of our simulation results is that, under the assumption
that environmental process do not strongly modify the dark matter distribution
(see PaperII), a firm detection of a
large core in any of the Milky Way satellites with a stellar mass
below few $10^6\Msun$ will call for a revision of the simple Cold
Dark Matter model . It will possibly point towards a different nature for dark
matter, either  warm \citep[but see][]{Maccio2012b}, or self
interacting \citep{Vogelsberger2014b,Elbert2015} or even more exotic
models \citep[e.g.][]{Maccio2015}.

\subsection{Diversity of star formation histories and the DM assembly}
\label{sec:SFH}

In this section we want to better understand the origin of the
diversity in star formation  histories shown in figure
\ref{fig:sfr}. To this extent, we will focus our attention on just four
haloes  that have very similar dark matter masses, all around
$10^{9.75} \Msun$, but have considerably   different stellar masses at
$z=1$ from $4.5 \times 10^4$ to $3.63 \times 10^6 \Msun$.

\begin{figure*}
\includegraphics[width=160mm]{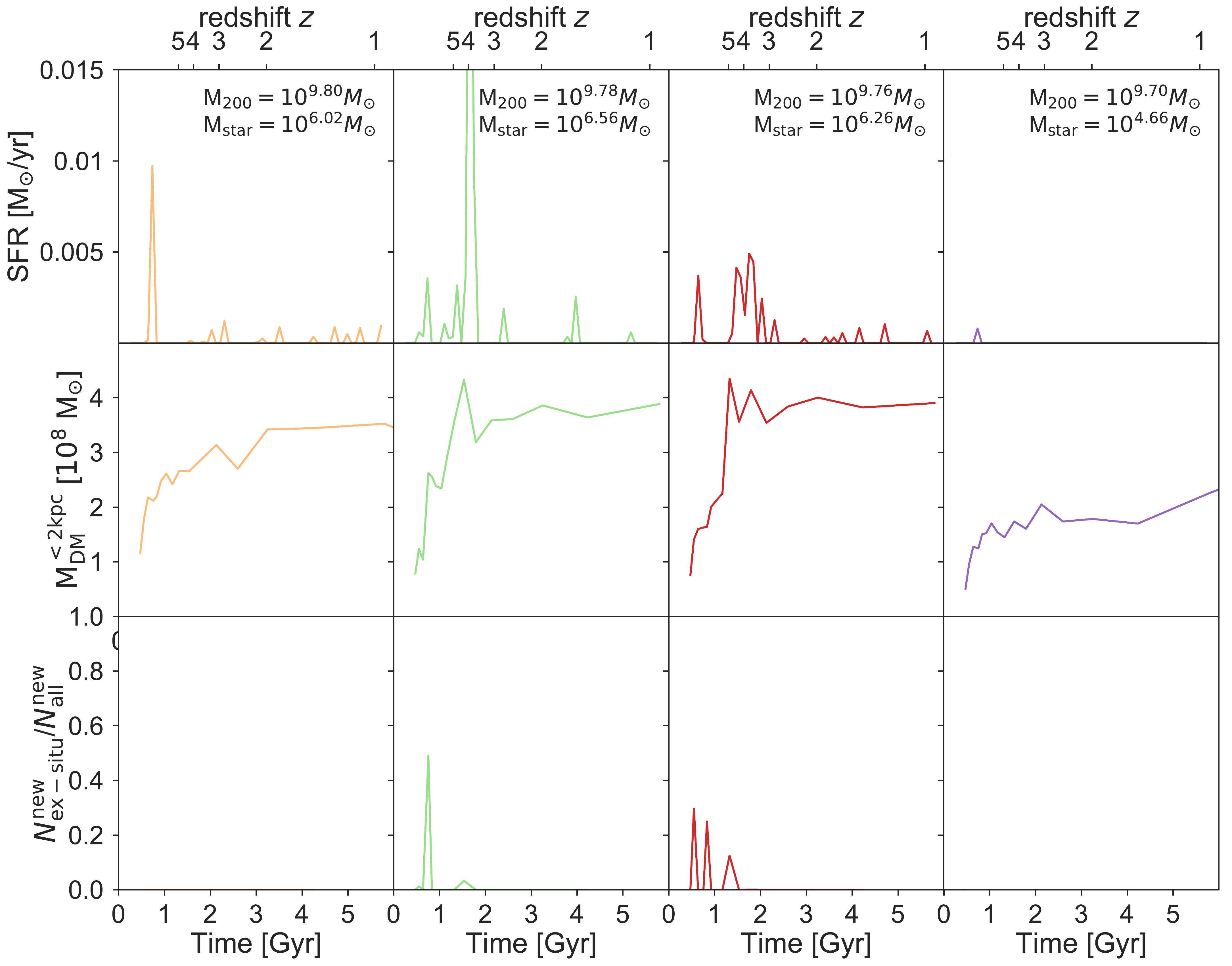}
\caption{The star formation rate (upper panels), the mass accretion
  history (middle panels) and the fraction of stars form from  in-situ vs. ex-situ gas particles (lower panesl). 
  The results are shown for four galaxies with  similar halo mass  at
  $z=1$, but very different stellar mass.  There is a clear
  correlation between mergers, i.e. sudden jump in the mass accretion
  history, and strong star formation episodes.}
\label{fig:mah2}
\end{figure*}

In  figure \ref{fig:mah2} we show the star formation histories of these
four galaxies (upper panels) compared to their inner mass accretion
(middle panels) defined as the mass  within a sphere of 2 kpc from the
center of the galaxy. 

There is a clear correlation between the infall of new mass (gas and
dark matter) and the onset of star formation. This is particularly
evident in the case of  sudden jumps in the enclosed mass, as for
example at $t\approx 1$ Gyr for the second (green) and the third (red)
object.  These are clearly major merger events that are able to double
the mass in  practically less than 100 Myrs, as we have also confirmed
by visual inspection. Corresponding to these mergers there is
a quick rise in star formation, which in the case of the ``red'' galaxy
is even followed by an extended ($\approx 2$ Gyr) period of activity. 

On the contrary, the fourth galaxy (purple) has an extremely quiet mass
accretion history,  characterized by continuous smooth accretion and no
mergers. In this case the star formation is limited to a single early
burst which led to a very low stellar mass at $z=1$. 

The increase in SFR is due to two effects: the merging halo brings in new gas
for star formation but it also compresses (due to shocks) the gas inside
the main halo, increasing its density and hence reducing the cooling time.

In order to disentangle these two effects, for every newly formed star
(during the SF peak) we traced back the origin of its parent gas particle and checked
if it was outside the virial radius at the previous snapshot (about 500 Myrs ago), meaning
that that gas particle came in inside the merging halo. 
The results are shown in the lowest panels of figure \ref{fig:mah2}.
For the two galaxies that undergone a merger, about half of the newly formed stars were 
generated from "ex-situ" gas particles. This results, even though based only on two galaxies, seem
to imply that there is an equal contribution of new and old gas to star formation.

Overall our analysis suggest  that the large scatter in the stellar mass
halo mass relation we find  at the ``edge'' of galaxy formation is due
to the strong impact that mergers have in triggering star formation
and to their intrinsically stochastic nature.

\subsection{Dark haloes}

A shown in figure \ref{fig:AM}, almost half of the haloes with a mass
below $5\times 10^9 \Msun$ did not manage to form any stars, and
remained dark. This is  due to the effect of the ultraviolet (UV)
background  parameterized following \cite{Haardt2001}. This
background takes into account the ionization field produced by quasars
and stars  and quenches star formation in small galaxies by
photo-heating their gas, which gets too hot to be confined in their
shallow potential wells \citep{Bullock2000, Somerville2002,Okamoto2008}.
The extent to which this field is able to affect star formation
depends on the halo mass, and it is usually described by the so called
characteristic mass ($M_c$) which is defined as the mass at which
haloes on average have lost half of their baryons \citep[e.g.][]{Simpson2013}

\begin{figure}
\includegraphics[width=85mm]{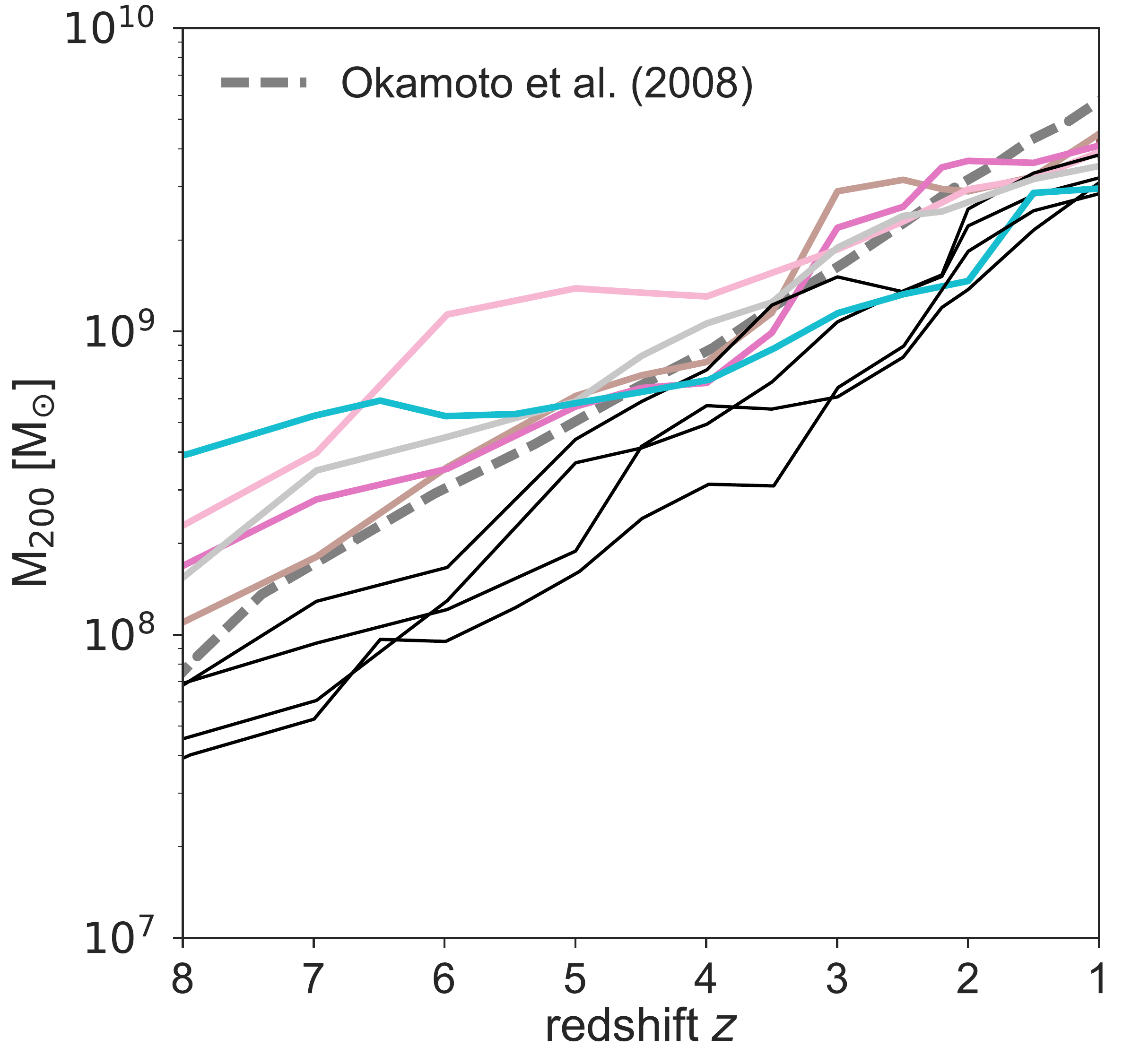}
\caption{ Mass accretion history of luminous and dark haloes with
  similar $z=1$ masses. The  grey dash line shows the characteristic
  mass $M_c$ (mass at which haloes on average have lost half of their
  baryons) as computed by Okamoto \etal 2008.}
\label{fig:dark}
\end{figure}

In figure \ref{fig:dark} following \cite{Fitts2016}, we show the mass accretion history of nine
haloes, four dark (black lines) and five luminous (colored lines), that have similar masses at
$z=1$, together with the redshift evolution of the characteristic mass
as computed by \citet[][dashed grey line]{Okamoto2008}. Luminous haloes have a more
rapid accretion history which brings their virial masses above $M_c$
at high redshift, allowing then gas to successfully cool in the
center of the halo. 
The three dark haloes, despite achieving the same final (z = 1) mass as the luminous ones, are characterized 
by a slow mass accretion at high redshift \citep{Benitez2017}.
This sets them below the critical mass at any redshift, and hence their gas cooling is strongly suppressed
in agreement with previous works \citep[][and references therein]{Sawala2016, Fitts2016}

\section{Discussion and Conclusions}
\label{sec:conclusions}

In this work we have presented a large set of cosmological
hydrodynamical simulations of the formation of several galaxies at the
lower ``edge'' of the galaxy mass spectrum.  We have simulated a total of 27
haloes with masses (at $z=1$) between $5\times 10^8<M_{\rm 200}<
2\times 10^{10}$ \Msun.  These simulations are aimed to characterize
the formation and evolution of todays galactic satellites {\it before}
they are accreted onto their parent halo. Using the zoom-in technique we
are able to attain a very high resolution both in mass (down to few
$10^2$ \Msun for gas) and space (with a softening of 20 pc).

Out of our 27 simulations, 19 end up with luminous haloes with stellar
masses between $2\times 10^4$ and $5\times 10^6 \Msun$, while eight do
not form any stars.

The luminous satellites successfully reproduce the main scaling
relations of todays Milky Way and M31 satellites, namely  the
size-velocity dispersion relation and the stellar mass metallicity
relation.  They have quite diverse star formation rates, ranging from
``extended'' bursty SF histories to a single star formation episode
in agreement with previous simulations \citep{Governato2015, Fitts2016}.
This large diversity generates a large scatter in the stellar mass -
halo mass relation. While the mean values are consistent with the
extrapolated results of $z=1$ abundance matching results from Moster
\etal 2013, the scatter, at a fixed halo mass can be as large as two
orders of magnitude, with a standard deviation of 0.45 dex.

By comparing the mass accretion and the star formation histories of
four haloes with the same final ($z=1$)  total halo mass (but
different stellar masses) we establish a clear correlation between the
large deviation in the final stellar mass  and the occurrence  (or
lack thereof) of substantial mergers events, which strongly impact the
star formation rate even in such small objects. Mergers have a double
effect, first to bring in new gas and second to compress and enhance
cooling of gas already in the halo. The intrinsic stochasticity of
halo mergers explains the large scatter in stellar mass at  a fixed halo
mass.

On the other hand eight haloes (out of 27) did not form any stars,
despite having at $z=1$ masses that are comparable if not higher than
luminous haloes. The presence of an uniform UV background is the main reason for the lack of gas cooling and hence star
formation in these haloes. We have shown that these haloes have a very
{\it slow} mass accretion history   which keeps them below the critical mass to retain
their baryons \citep{Gnedin2000,Hoeft2006,Okamoto2008}  and reach
high enough gas densities to have efficient cooling. 

Finally we look at the response of the dark matter distribution to
galaxy formation in our galaxies.  As in previous works
\citep{DiCintio2014a,Chan2015,Tollet2016} we find that the slope
of the DM profile ($\alpha$), calculated between 1-2\% of the
virial radius, correlates well with the efficiency of star formation,
defined as the ratio between stellar and halo mass.  When
compared with results from more massive galaxies from the NIHAO
simulation suite \citep{Wang2015} our galaxies seem to sit
on the same relation as their more massive counterparts.

Our results make the clear prediction that baryonic effects can be
neglected in isolated galaxies with stellar masses below few $10^6
\Msun$, on these scales the dark matter will retain its initial cuspy
slope as predicted from pure gravity simulations. 

This result confirm and extends previous findings from other
authors on slightly larger mass scales \cite[e.g.]{Governato2012, Onorbe2015}
and they implie that the unambiguous discovery of a cored dark
matter distribution in objects with stellar masses below this
threshold, will force us to rethink the nature of dark matter, since
in this case it will be very hard to  invoke a baryonic solution to
the problem.

Our paper covered only half of the life of these galaxies, from their
formation untill the supposed time of accretion that we fix to redshift
one. After being accreted,  galaxies will be subject to tidal forces,
ram pressure  and stripping. All these effects will contribute to
change some of the properties they had before accretion.  In the
companion paper (PaperII, Frings \etal 2017) we present an extensive
analysis of these environmental effects, which will give a
comprehensive picture of galaxy formation  and evolution at the edge
of its mass spectrum.

\section*{Acknowledgments} 

The authors would like to thank M. Collins and D. Weisz for proving their observational data in electronic form.
This research was carried out on the High Performance Computing resources at New York University Abu Dhabi; 
on the {\sc theo}  cluster of the Max-Planck-Institut f\"ur Astronomie and on the {\sc hydra}  clusters at the Rechenzentrum in Garching.
AVM and TB acknowledge funding from the Deutsche Forschungsgemeinschaft via the SFB 881 program 
"The Milky Way System" (subproject A2). JF and AVM acknowledge funding and support by the graduate college "Astrophysics of cosmological probes of gravity"
 by Landesgraduiertenakademie Baden-W\"uttemberg. AO acknowledges support from the German Science Foundation (DFG) grant 1507011 847150-0.
C. Penzo is supported by funding made available by ERC-StG/EDECS n. 279954

\vspace{-0.5cm}
\bibliographystyle{mnras}
\bibliography{bibliography}

\end{document}